\documentclass[useAMS,usenatbib]{mn2e}

\usepackage[]{txfonts}
\usepackage{subfigure}
\usepackage{epsfig}
\usepackage{url}
\def\simless{\mathbin{\lower 3pt\hbox
{$\rlap{\raise 5pt\hbox{$\char'074$}}\mathchar"7218$}}}   
\def\simmore{\mathbin{\lower 3pt\hbox
{$\rlap{\raise 5pt\hbox{$\char'076$}}\mathchar"7218$}}}   
\newcommand{\be}{\begin{equation}}
\newcommand{\ee}{\end{equation}}
\newcommand       \bea          {\begin{eqnarray}}
\newcommand       \eea          {\end{eqnarray}}
\newcommand       \apj          {ApJ}
\newcommand       \apjl         {ApJL}
\newcommand       \aap          {A\&A}
\newcommand       \nat          {Nature}
\newcommand       \mnras        {MNRAS}

\newcommand       \araa      {ARA\&A}

\newcommand      \apjs {ApJ Supplements}

\def\simlt{\mathrel{\hbox{\rlap{\hbox{\lower4pt\hbox{$\sim$}}}\hbox{$<$}}}}
\def\simgt{\mathrel{\hbox{\rlap{\hbox{\lower4pt\hbox{$\sim$}}}\hbox{$>$}}}}
\def\lesssim{\mathrel{\hbox{\rlap{\hbox{\lower4pt\hbox{$\sim$}}}\hbox{$<$}}}}

\def\rhoej{\rho_{\rm sh}}
\def\kabsnu{\kappa_{\rm abs,\nu}}
\def\kes{\kappa_{\rm es}}
\def\spr{s^{\prime}}
\def\fn{f_{\rm n}}

\def\sigmabf{\sigma_{\rm bf}}

\def\nA{n_{\rm A}}
\def\arec{\alpha_{\rm rec}}
\def\ne{n_{\rm e}}
\def\numin{\nu_{\rm min}}
\def\nutrh{\nu_{\rm thr}}

\def\dtrh{\Delta_{\rm thr}}

\def\gtrsim{\mathrel{\hbox{\rlap{\hbox{\lower4pt\hbox{$\sim$}}}\hbox{$>$}}}}

\title[Ionization Break-Out from Millisecond Pulsar Wind Nebulae]{Ionization Break-Out from Millisecond Pulsar Wind Nebulae: an X-ray Probe of the Origin of Superluminous Supernovae}

\author[]{Brian~D.~Metzger\thanks{E-mail: bmetzger@phys.columbia.edu}, Indrek~Vurm, Romain~Hasco\"{e}t, Andrei~M.~Beloborodov\\
Department of Physics and Columbia Astrophysics Laboratory, Columbia University, New York, NY, 10027}

\begin{document}
\date{Received / Accepted}
\pagerange{\pageref{firstpage}--\pageref{lastpage}} \pubyear{2012}

\maketitle

\label{firstpage}

\begin{abstract}

Magnetic spin-down of a rapidly rotating (millisecond) neutron star has been proposed as the power source of hydrogen-poor `superluminous' supernovae (SLSNe-I).  However, producing an unambiguous test that can distinguish this model from alternatives, such as circumstellar interaction, has proven challenging.  After the supernova explosion, the pulsar wind inflates a hot cavity behind the expanding stellar ejecta: the nascent millisecond pulsar wind nebula.  Electron/positron pairs injected by the wind cool through inverse Compton scattering and synchrotron emission, producing a pair cascade and hard X-ray spectrum inside the nebula.  These X-rays ionize the inner exposed side of the ejecta, driving an ionization front that propagates outwards with time.  Under some conditions this front can breach the ejecta surface within months after the optical supernova peak, allowing $\sim$ 0.1$-$1 keV photons to escape the nebula unattenuated with a characteristic luminosity $L_{\rm X} \sim 10^{43}-10^{45}$ erg s$^{-1}$.  This `ionization break-out' may explain the luminous X-ray emission observed from the transient SCP 06F, providing direct evidence that this SLSN was indeed engine-powered. Luminous break-out requires a low ejecta mass and that the spin-down time of the pulsar be comparable to the photon diffusion timescale at optical maximum, the latter condition being similar to that required for a supernova with a high optical fluence.  These relatively special requirements may explain why most SLSNe-I are not accompanied by detectable X-ray emission.  Global asymmetry of the supernova ejecta increases the likelihood of an early break-out along the direction of lowest density.  Atomic states with lower threshold energies are more readily ionized at earlier times near optical maximum, allowing `UV break-out' across a wider range of pulsar and ejecta properties than X-ray break-out, possibly contributing to the blue/UV colors of SLSNe-I.  
\end{abstract} 
  

\section{Introduction} 
\label{intro}

The recent discovery of ``superluminous" supernovae (SLSNe), with peak luminosities and radiated energies $\sim 10-100$ times higher than those of normal core collapse supernovae, shows that massive stars end their lives in a wider diversity of ways than previously anticipated \citep{Gal-Yam12}.  Some SLSNe, classified as Type IIn (``SLSN-II"), are likely powered by interaction with dense, hydrogen-rich circumstellar medium (e.g.~\citealt{Ofek+07}; \citealt{Smith+07,Smith+10}; \citealt{Metzger10}; \citealt{Rest+11}; \citealt{Moriya+13}; \citealt{Pan+13}).  This matter could be ejected in massive eruptions from the progenitor star that (for reasons unknown) occur just years prior to core collapse (e.g.~\citealt{Smith&Owocki06}).  

Other SLSNe are hydrogen-poor (``SLSN-I"), such as the class characterized by blue spectra with broad absorption lines of intermediate mass elements (\citealt{Quimby+07}; \citealt{Barbary+09}; \citealt{Pastorello+10}; \citealt{Chomiuk+11}; \citealt{Leloudas+12}; \citealt{Berger+12}; \citealt{Lunnan+13}; \citealt{Inserra+13}).  These events are associated with faint and metal poor galaxies with high star formation rates (\citealt{Quimby+11}; \citealt{Neill+11}; \citealt{Chomiuk+11}; \citealt{Chen+13}), apparently similar to the host galaxies of long duration gamma-ray bursts (however, see \citealt{Berger+12}, \citealt{Chornock+13}).  

The energetics and host environments of SLSN-I point towards the explosion of young, massive stars.  However, the ultimate source of their prodigious luminosities remains debated.  As in the case of the hydrogen-rich SLSNe, SLSN-I could be powered by circumstellar interaction with a {\it hydrogen-free} environment (\citealt{Blinnikov&Sorokina10}; \citealt{Chevalier&Irwin11}; \citealt{Ginzburg&Balberg12}; \citealt{Moriya&Maeda12}).  However, the lack of observed emission lines may be problematic for this scenario, since circumstellar material should in general extend across a range of radii, including optically-thin regions where shock heating should render such lines prominent.

\begin{figure*}
\includegraphics[width=1.0\textwidth]{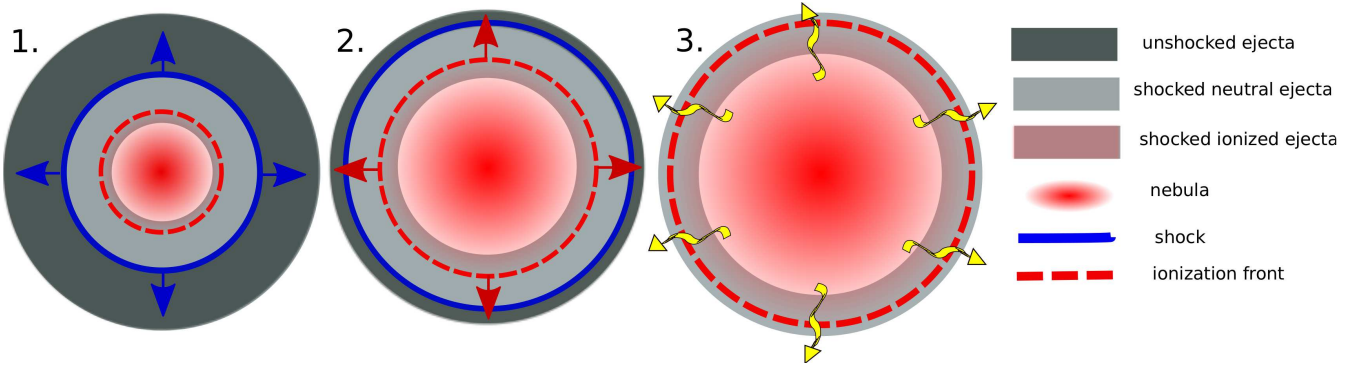}
\caption{Stages in ionization break-out from a young millisecond pulsar wind nebula: (1) Pulsar wind injects energetic pairs into the nebula, which cool rapidly, generating a pair cascade and hard X-ray radiation field.  The high pressure of the nebula drives a shock outwards through the supernova ejecta.  For sufficiently luminous pulsar winds, this shock reaches the ejecta surface within a few weeks after the initial explosion.  (2)  Nebular X-rays ionize the inner exposed side of the shocked ejecta, forming an ionization front that propagates outwards with time.  (3) Ionization front reaches the surface of the ejecta, allowing UV or X-ray photons to escape the nebula on the (short) electron scattering diffusion timescale.  The actual case is characterized by multiple ionization fronts, set by the elements and ionization states of relevance in different frequency ranges.  Break-out generally occurs progressively later for higher energy photons (Fig.~\ref{fig:layers}). } 
\label{fig:stages}
\end{figure*}

The high luminosities of SLSN-I could alternatively result from energy injection by a central engine, such as the magnetic spin-down of a newly-born rapidly-rotating neutron star (\citealt{Kasen&Bildsten10}; \citealt{Woosley10}; \citealt{Metzger+11}; \citealt{Dessart+12}) or accretion onto a newly-formed black hole (\citealt{Quataert&Kasen12}; \citealt{Dexter&Kasen12}).  SLSN-I light curves have been fit successfully to the neutron star model for initial rotation periods $P  \sim 1-8$ ms and surface dipole magnetic fields in the range $B_{\rm d} \sim 1-3\times 10^{14}$ G characteristic of Galactic `magnetars' (e.g.~\citealt{Chomiuk+11}; \citealt{Dessart+12}; \citealt{Inserra+13}).  We refrain from using the term magnetar in this work since comparably bright supernovae can in principle originate from millisecond neutron stars born with lower magnetic fields $B_{\rm d} \sim 10^{13}$ G more characteristic of the average pulsar population.  

Despite growing evidence in support of an engine-powered origin for some SLSNe, the millisecond pulsar model is challenging to test with confidence because the physically allowed range of $B_{\rm d}$ and $P$ can reproduce a wide range of peak luminosities and light curve durations.  Although the shape of the late time light curve can serve as a discriminant (\citealt{Inserra+13}), in practice it is difficult to rule out an important contribution from radioactive heating.  

Several theoretical issues furthermore remain unaddressed.  \citet{Kasen&Bildsten10}, for instance, assume that the nascent pulsar wind deposits thermal energy behind the supernova ejecta.  The resulting nebula sweeps up the ejecta into a thin shell and powers the supernova light curve via radiative diffusion.  The validity of this picture depends crucially on whether the Poynting flux of the wind is indeed converted into bulk kinetic energy, thus allowing a wind termination shock to form; this is the classical `$\sigma$ problem' of normal pulsar wind nebulae \citep{Kennel&Coroniti84}.  If such dissipation is not efficient, then a sufficiently strong toroidal magnetic field may accumulate to drive a bipolar jet (\citealt{Bucciantini+07,Bucciantini+08}).  Magnetic dissipation may occur due to MHD instabilities that develop inside the nebula (e.g.~\citealt{Begelman98}; \citealt{Porth&Komissarov13}), making it more likely that any jet will be stifled behind the supernova ejecta; however, this conversion is unlikely to be 100 percent efficient, and even a modest residual magnetization can be important dynamically (\citealt{Begelman&Li92}; \citealt{Bucciantini+07}).  Even if the pulsar energy is dissipated by a strong shock, precisely how the e$^{\pm}$ pairs injected by the pulsar thermalize their energy to be radiated by the ejecta  as an optical supernova remains unaddressed.

Given the degeneracies of the pulsar model and these nagging theoretical issues, it is important to develop alternative tests within a self-consistent frame-work.  In what follows we describe a potentially observable consequence of the pulsar model for VLSNe (other than its bright optical luminosity), which we justify more rigorously and develop in greater detail throughout the remainder of the paper.  

\subsection{Ionization Break-Out: Basic Physical Picture}

Young pulsar winds are composed primarily of electron/positron pairs\footnote{An important exception is the baryon-dominated neutrino wind that accompanies the first few minutes of the life of the neutron star (e.g.~\citealt{Qian&Woosley96}).} which are injected with a very high energy per particle, creating a nascent pulsar wind nebula (PWNe) behind the expanding supernova ejecta.  At early times this nebula is small and opaque, with a compactness parameter $\ell \gg 1$ for the millisecond pulsars of interest (eq.~[\ref{eq:compactness}]).  Newly-injected pairs rapidly lose their energy to inverse Compton scattering and synchrotron radiation.  This hard radiation in turn produces more pairs, resulting in a standard pair cascade and flat non-thermal X-ray spectrum $J_{\nu} \propto \nu^{-\alpha}$ with $\alpha \approx 1$ (e.g.~\citealt{Svensson87}). 

Normally in supernovae, the initially ionized ejecta recombines as it expands and cools.  The ejecta thus becomes more transparent to optical radiation with time (causing the supernova to brighten), but the bound-free opacity to soft X-rays remains enormous.  However, an important difference in the case of a sufficiently energetic pulsar wind is the high X-ray luminosity of the nebula, which photo-ionizes the inner exposed side of the ejecta.  As the ejecta expands, its density decreases, causing recombination to become less efficient.  This allows an ionization front(s) to form and propagate outwards with time.  If such a front reaches the surface of the ejecta, then high energy photons from the nebula are free to escape almost directly, producing bright X-ray emission starting several months after the explosion.  The stages in this process of `ionization break-out' are summarized in Figure \ref{fig:stages}.  

\begin{figure}
\includegraphics[width=0.5\textwidth]{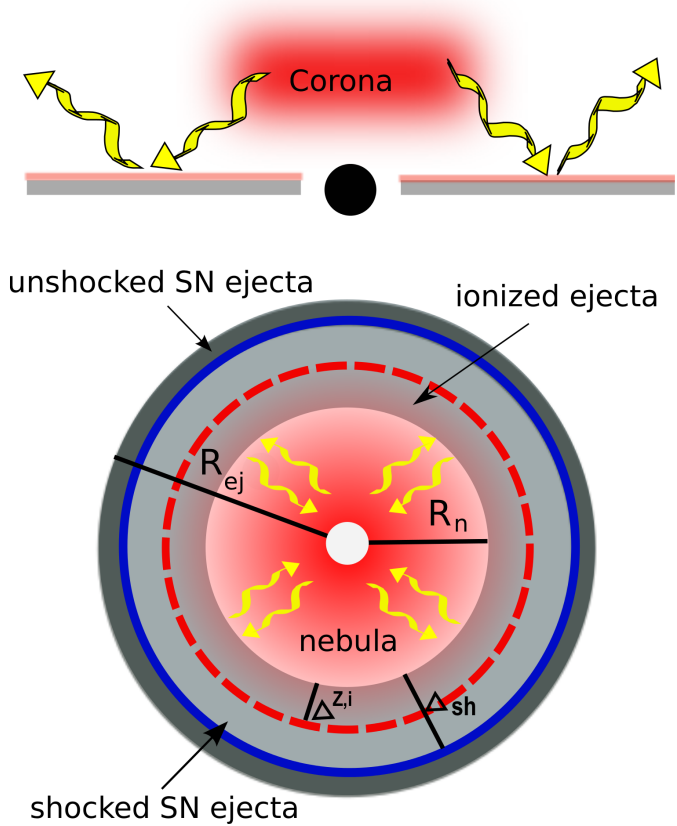}
\caption{The interaction between X-rays in a young PWNe with the expanding ejecta in an engine-powered supernova ({\it bottom}) mirrors the reprocessing of an X-ray corona by a black hole accretion disk ({\it top}), aside from the obvious difference between spherical and planar geometries and the multiple reflections experienced by photons in PWNe. } 
\label{fig:compare}
\end{figure}

Except for the difference in geometry, the interaction between nebular X-rays and the surrounding ejecta in an engine-driven supernova is similar to the reflection of coronal X-rays from the disk in an accreting black hole system (Figure \ref{fig:compare}).  X-rays that encounter the ejecta walls are either reflected or absorbed, depending on the albedo of the ionized layer separating the nebula from the bulk of the neutral ejecta.  The albedo in turn depends on the ionization state of the layer, which itself depends on the irradiating X-ray flux.  The characteristic size, X-ray luminosity, temperature, and ionization parameter characterizing millisecond PWNe are all in fact remarkably similar to those of AGN accretion disk systems (e.g.~\citealt{Lightman&Zdziarski87}; \citealt{Ross&Fabian93}; \citealt{Stern+95}; \citealt{Ross+99}; \citealt{Nayakshin&Kallman01}; \citealt{Kallman+04}).  One other difference between SLSNe and AGN is the temporal evolution, which is rapid in SLSNe due to the expanding ejecta, in contrast to the practically stationary geometry of AGN disks.

The expectation that bright X-ray emission should accompany some SLSN-I is of particular recent interest given the discovery of luminous soft X-ray emission ($L_{\rm X} \sim 10^{45}$ erg s$^{-1}$) from the superluminous supernova SCP 06F (\citealt{Barbary+09}; \citealt{Chatzopoulos+09}) roughly one month after the optical maximum (\citealt{Levan+13}).   One motivation of this work is to explore whether this emission is consistent with ionization break-out from a millisecond PWNe, as this could provide direct evidence that at least some SLSNe are engine-driven.

\subsection{This Paper}

Developing a fully self-consistent model for ionization break-out in SLSNe represents a formidable challenge.  In principle one must simultaneously follow the high energy radiative processes occurring inside the nebula; calculate the time-dependent, non-equilibrium ionization state of the ejecta; include the multi-frequency radiative transport responsible for exchanging energy between the nebula, ejecta, and external surroundings; and account for multi-dimensional effects, such as hydrodynamic instabilities (e.g.~Rayleigh-Taylor) and MHD processes (e.g.~magnetic dissipation and jet formation).   

In this paper, rather than tackle all of these issues in full detail, we instead present a simple model for the evolution of millisecond PWNe and their associated radiation that we believe captures the basic ingredients.  Our primary goal is to characterize the requisite conditions for ionization break-out as described above, and to determine the characteristic timescale, luminosity, and spectrum of the X-ray emission in those cases where break-out is indeed achieved.  In $\S\ref{sec:model}$ we describe the key components of the system.  In $\S\ref{sec:toy}$ we describe our approximate model for the evolution of the nebula and ejecta.  In $\S\ref{sec:results}$ we describe the results of our calculations.  In $\S\ref{sec:discussion}$ we discuss the implications of our results for X-ray ($\S\ref{sec:SLSNeX}$) and UV ($\S\ref{sec:UV}$) emission from SLSN-I.  We summarize our conclusions in $\S\ref{sec:conclusions}$.  Appendix \ref{sec:hlike} provides analytic estimates of the conditions required for the complete ionization of the ejecta in the simplified case of hydrogen-like atomic species.  Appendix \ref{sec:iondepth} provides an analytic estimate of the structure of the ionized layer and the frequency dependence of the photon penetration depth for a single ionization state.  Table \ref{table:defs} summarizes the definitions of frequently used variables.  

\section{PWNe Model}
\label{sec:model}

\begin{table}
\begin{scriptsize}
\begin{center}
\vspace{0.05 in}\caption{Definitions of Frequently Used Variables}
\label{table:defs}
\begin{tabular}{ll}
\hline \hline
\multicolumn{1}{c}{Variable} &
\multicolumn{1}{c}{Definition} \\
\hline
\\
$M_{\rm ej}$ & Total mass of the supernova ejecta, normalized as $M_3 = M_{\rm ej}/3M_{\odot}$\\
$v_{\rm ej}$ & Characteristic velocity of the ejecta, normalized as $v_9 = v_{\rm ej}/10^{9}$ cm s$^{-1}$\\
$R_{\rm ej}$ & Radius of outer edge of the ejecta \\
$V_{\rm ej}$ & Volume of the ejecta \\
$X_{A}$ & Mass fraction in the ejecta of element of atomic mass $A$ \\
$\rho_{\rm ej}$ & Initial density profile of the ejecta (eq.~[\ref{eq:rho0}])\\
 $\delta$ & Radial slope of the initial density profile of the ejecta \\
$P$ & Initial rotation period of the pulsar, normalized as $P_{-3}$ = $P$/1 ms \\
$E_{\rm rot}$ & Initial rotational energy of the pulsar \\
$B_{\rm d}$ & Dipole magnetic field of the pulsar, normalized as $B_{13} = B_{\rm d}/10^{13}$ G \\
$L_{\rm sd}$ & Spin-down luminosity of the pulsar, normalized as $L_{45} = L_{\rm sd}/10^{45
}$ erg s$^{-1}$ (eq.~[\ref{eq:Lsd}])\\
$t_{\rm sd}$ & Initial spin-down time of the pulsar (eq.~[\ref{eq:tsd}]) \\
$t_{\rm d}$ & Photon diffusion time of the ejecta (eq.~[\ref{eq:td2}]) \\
$t_{\rm d,0}$ & Photon diffusion timescale of the ejecta at optical peak (eq.~[\ref{eq:td}])\\
$R_{\rm n}$ & Radius of the nebula \\
$\ell$ & Compactness parameter of the nebula to non-thermal radiation (eq.~[\ref{eq:compactness}]) \\
$V_{\rm n}$ & Volume of the nebula \\
$f_{\rm sh}$ & Shocked fraction of the ejecta (eq.~[\ref{eq:fsh}]) \\
$R_{\rm cr}$ & Radius at which the shock reaches the front of the ejecta \\
$B_{\rm n}$ & Characteristic magnetic field strength in the nebula (eq.~[\ref{eq:Bn}]) \\
$E$ & Total energy of the nebula and ejecta in radiation (eq.~[\ref{eq:dEdt}]) \\
$E_{\rm th}$ & Total thermal energy of the nebula and ejecta (eq.~[\ref{eq:thermo_evo}]) \\
$E_{\rm nth,\nu}$ & Spectral energy distribution of non-thermal photons in the nebula \\
$E_{\rm nth}$ & Total energy of non-thermal photons in the nebula (eq.~[\ref{eq:photon_evo}]) \\
$\Delta_{\rm sh}$ & Thickness of the shocked layer of swept-up ejecta (eq.~[\ref{eq:delta_sh}]) \\
$\Delta^{Z,i}$ & Thickness of the ionized layer of element $Z$, ionization state $i$ (eq.~[\ref{eq:ionthick}]) \\
$\nu_{\rm thr}^{Z,i}$ & Ionization threshold frequency of atomic species $Z$, ionization state $i$ \\
$f_{\rm n}^{Z,i}$ & Neutral fraction in ejecta of atomic species $Z$, ionization state $i$ \\
$\kappa_{\rm bf,\nu}^{Z,i}$ & Bound-free absorption opacity in ejecta due to atomic species $Z$, ionization state $i$ \\
$\mathcal{A}_{\nu}$ & Albedo of the layer of ionized ejecta to nebular photons of frequency $\nu$ \\
$\xi$ & Ionization parameter of nebular radiation interacting with ejecta walls (eq.~[\ref{eq:ion_param}]) \\
$T_{\rm C}$ & Compton temperature of electrons (eq.~[\ref{eq:Tcompton}]) \\
$ \tau_{\rm es}^{\rm n}$ & Electron scattering optical depth through the nebula (eq.~[\ref{eq:taueseq}]) \\
$\tau_{\rm es}^{\rm ej}$ & Electron scattering optical depth through the ejecta (eq.~[\ref{eq:tau_ej}])\\
$\tau_{\rm es}^{\rm sh}$ & Electron scattering optical depth through the shocked ejecta (eq.~[\ref{eq:tauessh}])\\
$L_{\rm SN}$ & Luminosity of the supernova (eq.~[\ref{eq:L_SN}]) \\
$t_{\rm d}^{\rm n}$ & Photon diffusion time through the nebula (eq.~[\ref{eq:tdiffn}]) \\
$J_{\nu}$ & Mean specific intensity of photons in nebula (eq.~[\ref{eq:Jnu}]) \\
$T_{\rm th}$ & Blackbody temperature of the nebula and ejecta (eq.~[\ref{eq:Tn}]) \\
$\eta$ & Ratio of absorption to scattering opacity in the ionized layer at frequency $\nu$ (eq.~[\ref{eq:eta1}]) \\
$\eta_{\rm thr}$ & Ratio of absorption to scattering opacity in the ionized layer at frequency $\nu_{\rm thr}^{Z,i}$ (eq.~[\ref{eq:eta}]) \\
$t_{\rm bo}$ & Time of `ionization break-out' at UV or X-ray wavelengths (eq.~[\ref{eq:tbo}]) \\
$L_{\rm bo,X}$ & X-ray luminosity ($\sim 0.1-1$ keV) at ionization break-out (eq.~[\ref{eq:LXbo}])\\
 \\
  \\
\hline
\hline

\end{tabular}
\end{center}
\end{scriptsize}
\end{table}


Models for PWNe and their radiation have been developed previously by many authors (\citealt{Chevalier&Fransson92}; \citealt{vanderSwaluw+04}; \citealt{Gaensler&Slane06}; \citealt{Gelfand+09}; \citealt{Martin+12}; \citealt{Kotera+13}).  Little work, however, has been dedicated to strongly magnetized millisecond pulsars, especially at the earliest stages of evolution.  One exception is the recent work by \citet{Kotera+13}, to which we compare our results in $\S\ref{sec:conclusions}$.  We will show that the evolution of millisecond PWNe can differ qualitatively from those of less energetic pulsars due to the influence of nebular radiation on the X-ray opacity of the ejecta.  This can produce an early stage of ionization and UV/X-ray transparency that does not occur for pulsars with lower spin-down luminosities.

Figure \ref{fig:compare} shows a schematic diagram of the PWN system.  At any time, the system is composed of several distinct regions: (1) the unshocked ejecta, of outer radius $R_{\rm ej}$ ($\S\ref{sec:ejecta}$) and interior volume $V_{\rm ej} = 4\pi R_{\rm ej}^{3}/3$; (2) the nebula, of radius $R_{\rm n} \lesssim R_{\rm ej}$ and volume $V_{\rm n} = 4\pi R_{\rm n}^{3}/3$ ($\S\ref{sec:pulsar}$); and (3) an inner shell of ejecta, of radius $\sim R_{\rm n}$ and thickness $\Delta_{\rm sh} \sim R_{\rm n}/10$, compressed by a shock driven outwards by the high nebular pressure ($\S\ref{sec:size}$).  The dissipative nebula is a slowly expanding bubble filled with radiation and $e^{\pm}$ pairs which are cooled to non-relativistic temperature.  Radiation strongly dominates the internal energy and pressure in all three regions.  

The radiation field of the system can be divided into a thermal bath with energy $E_{\rm th}$ and blackbody temperature
\be
T_{\rm th} = \left(\frac{E_{\rm th}}{aV_{\rm ej}}\right)^{1/4},
\label{eq:Tn}
\ee
and (2) non-thermal radiation with total energy $E_{\rm nth} = \int E_{\rm nth,\nu}d\nu$ and spectrum $E_{\rm nth,\nu}$.  Radiation dominates the total energy of the system $E = E_{\rm nth} + E_{\rm th}$.  Non-thermal photons irradiate the inner exposed side of the ejecta, producing an ionization front of depth $\Delta^{Z,i}$ ($\S\ref{sec:opacity}$) for each element $Z$ and ionization state $i$ of relevance.

\subsection{Supernova Ejecta}
\label{sec:ejecta}

We assume that the core collapse mechanism ejects an envelope of mass $M_{\rm ej} = 3M_3 M_{\odot}$ with a velocity $v_{\rm ej} = v_9 10^{9}$ cm s$^{-1}$ from a star of initial radius $R_\star \sim 10R_{\odot}$ characteristic of the compact stars of interest.  Fits to the light curves of SLSN-I provide characteristic values $M_3 \sim 1-3$ and $v_9 \sim 1-2$ (e.g.~\citealt{Inserra+13}), corresponding to kinetic energies $M_{\rm ej}v_{\rm ej}^{2}/2 \approx 10^{51}-10^{52}$ ergs.

 Within a few expansion times, $\sim R_\star/v_{\rm ej} \sim 10^{3}$ s, the ejecta enters self-similar evolution.  The density structure of the exploded star (for $r < R_{\rm ej}$) is given by (\citealt{Kasen&Bildsten10})
\be
\rho_{\rm ej}(r,t) = \frac{(3-\delta)}{4\pi}\frac{M_{\rm ej}}{R_{\rm ej}(t)^{3}}\left(\frac{r}{R_{\rm ej}(t)}\right)^{-\delta},
\label{eq:rho0}
\ee
where $R_{\rm ej} = v_{\rm ej}t$ is the characteristic outer radius of the ejecta, and we usually take $\delta = 0-1$.\\

The ejecta is composed primarily of elements with charge $Z \ge 2$ and mass $A \simeq 2Z$, such as $^{4}$He, $^{12}$C, $^{16}$O, $^{20}$Ne, $^{24}$Mg, $^{28}$Si, $^{32}$S, and $^{56}$Fe.  Lacking a detailed model for the progenitor structure and resulting explosive nucleosynthesis, we take as representative the mass fractions $X_A$ from the 10 $M_{\odot}$ He core model of \citet{Nakamura+01} for a supernova energy $E_{\rm SN} = 10^{52}$ ergs: $X_{\rm He} = 0.15; X_{\rm C} = 0.01, X_{\rm O} = 0.59, X_{\rm Ne} = 0.04, X_{\rm Mg} = 0.02, X_{\rm Si} = 0.03, X_{\rm S} = 0.01, X_{\rm Fe} = 0.05$.  An oxygen-dominated composition appears to be a robust feature of the energetic explosion of a compact star (e.g.~\citealt{Maeda+02}).  For simplicity we assume that the elements are  mixed homogeneously throughout the ejecta, as may be justified physically by the effects of Rayleigh-Taylor instabilities at the contact discontinuity separating the nebula from the shocked ejecta (e.g.~\citealt{Blondin+01}; \citealt{Hammer+10}).  

Radiation is free to escape the nebula once the photon diffusion timescale $t_{\rm d} \sim 3M_{\rm ej}\kappa/4\pi R_{\rm ej}c$ becomes less than the expansion timescale of the ejecta $\sim R_{\rm ej}/v_{\rm ej}$, where $\kappa$ is the opacity \citep{Arnett82}.  This occurs after a characteristic time
\be
t_{\rm d,0} = \left(\frac{3}{4\pi}\frac{\kappa M_{\rm ej}}{v_{\rm ej}c}\right)^{1/2} \simeq 3\times 10^{6}M_{3}^{1/2}(\kappa/\kappa_{\rm es})^{1/2}v_9^{-1/2}\,{\rm s},
\label{eq:td}
\ee
where $\kappa$ is normalized to the electron scattering opacity for fully ionized matter, $\kappa_{\rm es} = 0.2$ cm$^{2}$ g$^{-1}$, which characterizes the escape of thermal radiation at optical wavelengths.   Hereafter we normalize times to $t_{\rm d,0}$ (for $\kappa = \kappa_{\rm es}$) since this sets the characteristic timescale for the optical luminosity to peak and the minimum timescale for any ionization front to reach the ejecta surface.       

\subsection{Pulsar Wind Nebula}
\label{sec:pulsar}
A neutron star born with an initial rotational period $P = 2\pi/\Omega = P_{-3}$ ms has an associated energy
\be
E_{\rm rot} = \frac{1}{2}I\Omega^{2} \simeq 2\times 10^{52}P_{-3}^{-2}{\rm\,ergs},
\label{eq:Erot}
\ee 
where $I \simeq 10^{45}$ g cm$^{2}$ is the neutron star moment of inertia.  The pulsar injects energy behind the supernova shock at a rate, which for an aligned force-free wind is given by
\begin{eqnarray}
L_{\rm sd} = \frac{\mu^{2}\Omega^{4}}{c^{3}} \simeq 6\times 10^{45}B_{13}^{2}P_{-3}^{-4}\left(1 + \frac{t}{t_{\rm sd}}\right)^{-2}{\rm erg\,s^{-1}}
\label{eq:Lsd}
\end{eqnarray}
where $\mu = B_{\rm d}R_{\rm NS}^{3}$ is the dipole moment; $B_{\rm d} = 10^{13}B_{13}$ G is the surface equatorial dipole field;\footnote{Note that our definition of $B_{\rm d}$ is a factor of $\sqrt{12} \simeq 3.5$ lower than that defined by \citet{Kasen&Bildsten10}, who instead assume vacuum dipole spin-down for an inclination angle of $\alpha = 45^{\circ}$ between the magnetic dipole and rotation axis.} $R_{\rm NS} = 10$ km is the neutron star radius; and
\be
t_{\rm sd} = \left.\frac{E_{\rm rot}}{L_{\rm sd}}\right|_{t = 0}\simeq 3\times 10^{6}B_{13}^{-2}P_{-3}^{2}\,{\rm s}
\label{eq:tsd}
\ee
is the initial spin-down time.  This luminosity inflates a nebula of relativistic particles and radiation inside the cavity evacuated by the expanding supernova ejecta (Figure \ref{fig:compare}).  Powering a SLSN  of peak luminosity $L_{\rm p} \gtrsim 10^{44}$ erg s$^{-1}$ via pulsar spin-down requires that $L_{\rm sd} \gtrsim L_{\rm p}$ and that $t_{\rm sd}$ be comparable to, or somewhat shorter than, $t_{\rm d,0}$ (eq.~[\ref{eq:td}]).  These conditions can be satisfied for $B_{13} \sim 0.3-30$ and $P_{-3} \sim 1-3$. 

Pulsar winds are composed primarily of $e^{\pm}$ pairs, which are injected into the nebula at a characteristic rate 
\be
\dot{N}_{\pm} = \mu_{\pm}\dot{N}_{\rm GJ} \simeq 5\times 10^{37}\mu_{\pm}B_{13}P_{-3}^{-2}(1 + t/t_{\rm sd})^{-1}{\rm s^{-1}},
\label{eq:dotNpm}
\ee
where $\dot{N}_{\rm GJ} \equiv 2B_{\rm d}\Omega^{2}R_{\rm NS}^{3}/e c$ is the \citet{Goldreich&Julian69} flux.  Here $\mu_{\pm}$ is the pair multiplicity, which is uncertain but is unlikely to exceed $\sim 10^{5}$ and could be much less in the case of magnetar-strength fields (e.g.~\citealt{Arons07}; \citealt{Medin&Lai10}; \citealt{Beloborodov13}).  If the initial Poynting flux of the pulsar wind is converted to random thermal energy of the pairs (through shocks or magnetic reconnection), then the maximum energy per particle is
\be
\epsilon_{\pm} = \frac{L_{\rm sd}}{\dot{N}_{\pm}} \simeq 1.1\times 10^{8}{\rm \,erg\,\,} \mu_{\pm}^{-1}B_{13}P_{-3}^{-2} \left(1 + t/t_{\rm sd}\right)^{-1},
\label{eq:epsilonpm}
\ee
corresponding to a random particle Lorentz factor
\be
\gamma_{\pm} = \frac{\epsilon_{\pm}}{m_e c^{2}} \simeq 1.3\times 10^{14}\mu_{\pm}^{-1}B_{13}P_{-3}^{-2}\left(1 + t/t_{\rm sd}\right)^{-1}.
\ee
In practice such a large particle energy is unlikely to be achieved because cooling truncates the particle acceleration process.  Shock acceleration cannot, for instance, occur on a timescale faster than the gyration period of the accelerated pair $t_{\rm g} = 2\pi \gamma_{\pm}m_e c/e B_{\rm n}$ in the magnetic field of the nebula $B_{\rm n}$.  Equating this to the synchrotron cooling timescale $t_{\rm syn} = 6\pi m_e c/\sigma_{\rm T}B_{\rm n}^{2}\gamma_{\pm}$ limits the pair Lorentz factor to a maximum value
\be
\gamma_{\pm}^{\rm max} = \sqrt{\frac{3e}{\sigma_{\rm T}B_{\rm n}}} \simeq 5\times 10^{6}\epsilon_{\rm B,-2}^{-1/4}L_{45}^{-1/4}v_{9}^{1/2}M_{3}^{1/4}\left(\frac{t}{t_{\rm d,0}}\right)^{1/2},
\label{eq:gamma_max}
\ee
where $\sigma_{\rm T}$ is the Thomson cross section and $L_{45} \equiv L_{\rm sd}/10^{45}$ erg s$^{-1}$. The magnetic field in the nebula
\be
B_{\rm n} \approx 80\,{\rm G}\,\epsilon_{\rm B,-2}^{1/2}L_{45}^{1/2}v_{9}^{-1}M_{3}^{-1/2}\left(\frac{t}{t_{\rm d,0}}\right)^{-1},\,\,\,\,\
\label{eq:Bn}
\ee
is estimated by assuming that the magnetic energy $B_{\rm n}^{2}V_{\rm n}/8\pi$ of the nebula is a fraction $\epsilon_{\rm B} = 10^{-2}\epsilon_{\rm B,-2}$ of the total energy injected by the pulsar $\sim L_{\rm sd}t$ over time $\sim t$, where $V_{\rm n} \simeq 4\pi R_{\rm n}^{3}/3 \sim 4\pi R_{\rm ej}^{3}/3$ is the nebula volume.  Small values of $\epsilon_{B} \lesssim 0.01$  are motivated by the likelihood of magnetic dissipation due to instabilities in the nebula (e.g.~\citealt{Porth&Komissarov13}) and our assumption that no jet forms.  

An upper limit on the pair Lorentz factor similar to equation (\ref{eq:gamma_max}) results if cooling is instead dominated by IC scattering.  Whether IC or synchrotron dominates cooling depends on the value of $\epsilon_{\rm B}$, the energy density of the thermal bath, and Klein-Nishina corrections.  

\subsection{Pair Cascade}
\label{sec:cascade}

Regardless of what process dominates pair cooling, pair acceleration is accompanied by additional pair production.  A pair Lorentz factor $\gamma_{\pm} \sim 10^{3} \ll \gamma_{\pm}^{\rm max}$ is sufficient to up-scatter seed thermal photons from the nebula ($h \nu \gtrsim $ 1 eV) to energies $\gtrsim 2 m_e c^{2} \approx$ 1 MeV.  The maximum energy of synchrotron photons $h\nu_{\rm syn} = \hbar eB_{\rm n}(\gamma_{\pm}^{\rm max})^{2}/m_e c \sim 100\,{\rm MeV}$ also exceeds the pair creation threshold. 

Pair creation is not only likely but copious due to the high `compactness' of the nebula,
\begin{eqnarray}
\ell &\equiv& \frac{E_{\rm nth}\sigma_{\rm T} R_{\rm n}}{V_{\rm n}m_e c^{2}} \sim \frac{3}{4\pi(1-\bar{\mathcal{A}}_{\nu})}\frac{\sigma_{\rm T}}{m_e c^{3} v_{\rm ej}}\frac{L_{\rm sd}}{t} \noindent \\
&\approx& 2.1(1-\bar{\mathcal{A}}_{\nu})^{-1}L_{45}v_9^{-1/2}M_3^{-1/2}\left(\frac{t}{t_{\rm d,0}}\right)^{-1},
\label{eq:compactness}
\end{eqnarray}
where the non-thermal energy of the nebula $E_{\rm nth}$ is approximated in the second equality using its value obtained by balancing energy injection from the pulsar with absorption by the ejecta walls,
\be
E_{\rm nth} \approx \frac{L_{\rm sd}}{1 - \bar{\mathcal{A}}_{\nu}}\frac{R_{\rm ej}}{c},
\label{eq:Ent_eq}
\ee 
as described in Appendix \ref{sec:hlike} (eq.~[\ref{eq:edotsdeq}]), where $\bar{\mathcal{A}}_{\nu} \lesssim 1$ is the albedo of the walls, appropriately averaged over frequency $\nu$.  

If $\ell \gg 1$ then most photons generated by the cooling pairs produce additional pairs.  These in turn up-scatter (or emit by synchrotron radiation) additional photons of sufficient energy to generate further pairs, and so on.  The result is a standard non-thermal `pair cascade' (e.g.~\citealt{Svensson87}; \citealt{Lightman&Zdziarski87}; \citealt{Vurm&Poutanen09}).

Pairs leaving the cascade cool rapidly to sub-relativistic temperatures.  The ratio of the Compton cooling timescale $t_{\rm C}$ for non-relativistic particles to the evolution timescale is given by
\be
\frac{t_{\rm C}}{t} = \frac{3 m_e c}{4\sigma_{\rm T}U_{\gamma}}\frac{v_{\rm ej}}{R_{\rm n}} \sim \frac{\pi m_e c v_{\rm ej}^{3}}{\sigma_{\rm T} L_{\rm sd}}t \approx 4\times 10^{-4}L_{45}^{-1}M_3^{1/2}v_9^{5/2}\left(\frac{t}{t_{\rm d,0}}\right) ,
\ee
where $U_{\gamma} = E/V_{\rm n}$ is the radiation energy density of the nebula, which is estimated in the second equality as $\sim L_{\rm sd}t/V_{\rm n}$.  As pairs cool, their energy$\mbox{--}$and, in effect, the entire spin-down luminosity of the pulsar$\mbox{--}$is thus distributed across a power-law photon spectrum $J_{\nu} \propto \nu^{-\alpha}$ ($\alpha \approx 1$), extending from the thermal bath $h\nu \sim $ 1 eV up to the pair threshold cut-off $h\nu \sim $ 1-10 MeV set by the pair cascade.  

The pair cascade converts a sizable fraction of the injected energy into pairs, resulting in an {\it effective} pair production rate
\be
\dot{N}_{\pm}^{\rm +} \simeq \frac{Y L_{\rm sd}}{m_e c^{2}} = 1.2\times 10^{50}\left(\frac{Y}{0.1}\right)L_{45}{\,\rm s^{-1}}
\ee
significantly higher than that initially injected by the pulsar (eq.~[\ref{eq:dotNpm}]), where $Y$ is the pair yield factor normalized to its characteristic value for a saturated cascade (\citealt{Svensson87}).  

Pairs leaving the cascade cool to an equilibrium temperature $T_{\pm} = T_{\rm C}$ set by the balance between Compton heating and cooling, as determined by the approximate requirement that  (e.g.~\citealt{Beloborodov&Illarionov95})
\be
\int J_{\nu}\left(h\nu - 4 k T_{\rm C} - \frac{21 (h\nu)^{2}}{5 m_e c^{2}}\right)d\nu = 0,
\label{eq:Tcompton}
\ee
where 
\be
J_{\nu} = \frac{c E_{\rm nth, \nu}}{4\pi V_{\rm n}}
\label{eq:Jnu}
\ee
is the mean specific intensity of photons in the nebula.  The Compton temperature of the nebula $k T_{\rm C}$ is typically $\sim 1$ keV.

Pair creation is offset by pair annihilation, which occurs at a characteristic rate
\be
\dot{N}_{\pm}^{\rm -} = \frac{3}{16}\sigma_{\rm T} c n_{\pm}^{2}V_{\rm n},
\ee
where $n_{\pm}$ is the pair density of the nebula.  The balance between pair creation ($\dot{N}_{\pm}^{\rm +}$) and annihilation ($\dot{N}_{\pm}^{-}$) results in a scattering optical depth through the nebula due to pairs given by
\begin{eqnarray}
\tau_{\rm es}^{\rm n} &=& \sigma_{\rm T} R_{\rm n}n_{\pm} = R_{\rm n}\left[\frac{16 \dot{N}_{\pm}^{-}\sigma_{\rm T}}{3  V_n c}\right]^{1/2} 
= \left[\frac{4 Y\sigma_{\rm T} L_{\rm sd}}{\pi R_n m_e c^{3}}\right]^{1/2} \nonumber \\ &\approx& 1.1\left(\frac{Y}{0.1}\right)^{1/2}L_{45}^{1/2}M_{3}^{-1/4}v_9^{-1/4}\left(\frac{t}{t_{\rm d,0}}\right)^{-1/2}\,\,\,\,({\rm equilibrium}),
\label{eq:taueseq}
\end{eqnarray} 
where we again have assumed $R_{\rm n} \sim R_{\rm ej}$ in the numerical estimate.  The timescale to reach this equilibrium, $t_{\rm eq} \simeq 16 R_{\rm n}/3 c \tau_{\rm es}^{\rm n}$, is short compared to the evolution timescale $t = R_{\rm ej}/v_{\rm ej}$ as long as
\be
\tau_{\rm es}^{\rm n} \gtrsim \frac{16}{3}\frac{v_{\rm ej}}{c} \simeq 0.18v_9
\ee
as is satisfied at times
\be
t \lesssim  1\times 10^{8}\left(\frac{Y}{0.1}\right)L_{45}v_9^{-3}{\rm s}.
\ee
Pair creation/destruction equilibrium is thus maintained on a timescale that is generally longer than the pulsar spin-down timescale $t_{\rm sd}$ (eq.~[\ref{eq:tsd}]).  The low value of $\tau_{\rm es}^{\rm n}$ implies that photons diffuse across the nebula on a timescale that is always shorter than that required to diffuse outwards into the surrounding ejecta. 

The flat non-thermal spectrum injected into the nebula by the cooling pairs is only maintained if hard photons are not thermalized by their interaction with the  background pairs.  A sufficiently high pair abundance would drive the photons into a relatively narrow Wien spectrum with a common temperature $T = T_{\pm} = T_{\rm C}$.  The ratio of the timescale for photon-pair equilibration $t_{\rm e-\gamma}$ to the evolution timescale can be written
\be
\frac{t_{\rm e-\gamma}}{t} = \frac{t_{\rm C}}{t}\frac{U_{\gamma}}{3n_{\pm} kT_{\pm}/2} = y^{-1}\frac{v_{\rm ej}}{c}
\ee
where
\be
y = \frac{4 kT_{\pm}\tau_{\rm es}^{\rm n}}{m_e c^{2}}
\ee
is the Compton y-parameter.  The low values of $k T_{\pm} = T_{\rm C} \lesssim 1 \,\,{\rm keV} \ll m_e c^{2}$ and of $\tau_{\rm es}^{\rm n} \lesssim$ few (eq.~[\ref{eq:taueseq}]) at times near or after the optical peak $t \sim t_{\rm d,0}$ imply that $y < \frac{v_{\rm ej}}{c}\sim 0.03 v_9$ and hence $t_{\rm e-\gamma} > t$.  We thus conclude that nebular photons cannot equilibrate with the pairs due to the higher heat capacity of photons in the radiation-dominated nebula ($U_{\gamma} \gg 3n_{\pm} kT_{\pm}/2$).  As we discuss below, absorption by the walls of the ejecta provides a more effective means to thermalize the radiation field.  

\subsection{Ejecta Opacity and Ionized Layer}
\label{sec:opacity}

The opacity of the ejecta is important both because it determines when the ejecta becomes transparent to radiation, and because it controls the albedo of the ejecta walls.  Several processes in principle contribute to the ejecta opacity 
\be
\kappa_{\nu} = \kappa_{\rm es} + \kappa_{\rm bb,\nu} + \kappa_{\rm ff,\nu} + \kappa_{\rm bf,\nu},
\ee
including electron scattering (es), bound-bound [lines] (bb), free-free (ff), and bound-free (bf) absorption.    

Free-free absorption is only important at low frequencies, below the optical band.  The dominant opacity at optical/UV wavelengths is (elastic) electron scattering and bound-bound absorption by Doppler-broadened atomic lines.  At early times the ejecta is sufficiently hot to be thermally ionized, while at later times photoionization by nebular X-rays maintains a reasonably high density of free electrons (we show later that helium and valence electrons of heavier elements are relatively easy to photo-ionize throughout the ejecta).  The bound-bound opacity is probably similar to that in SN Ia, $\kappa_{\rm bb} \sim 0.1$ cm$^{2}$ g$^{-1}$ (\citealt{Pinto&Eastman00}), although this may be an overestimate in our case due to the lower iron abundance in core collapse SN ($X_{\rm Fe} \approx 0.05$) and since the valence electrons of iron are relatively easily photo-ionized throughout the ejecta.  Since the ejecta temperature at times near the peak of the supernova emission typically corresponds to optical frequencies, the above discussion motivates using a constant opacity $\kappa \sim \kappa_{\rm es} = 0.2$ cm$^{2}$ g$^{-1}$ to calculate the cooling rate of the ejecta ($\S\ref{sec:energy}$).

At hard UV$/$soft X-ray wavelengths the dominant opacity is bound-free absorption, which depends sensitively on the ionization state of the ejecta.  Normally, supernova ejecta recombines as it expands and cools, eventually becoming largely neutral.  This results in an enormous bound-free opacity (several orders of magnitude higher than $\kappa_{\rm es}$), preventing soft X-rays from escaping on timescales of relevance.  In the case of energetic PWNe, however, the ionization state of the ejecta, and hence its X-ray opacity, is controlled by the balance between photo-ionization from nebular X-rays and radiative recombination.  

UV/soft X-ray photons penetrate to a depth $\Delta^{Z,i}$ through the ejecta (Fig.~\ref{fig:compare}) set by the location of the ionization front of the dominant element $Z$ and ionization state $i$ in the frequency range of interest.  The dominant elements are usually those with the highest abundance in the ejecta ($\S\ref{sec:ejecta}$), such as helium (UV), oxygen ($\lesssim$ keV  X-rays), and iron ($\lesssim 10$ keV X-rays).
Because the ionization depth $\Delta^{Z,i}$ is usually smaller for higher ionization states with larger threshold frequencies $\nu_{\rm thr}^{Z,i}$, the penetration depth of a photon is usually limited (ultimately) by the atomic species/ionization state with the closest value of $\nu_{\rm thr}^{Z,i}$ just below the photon frequency $\nu$.  

We define the penetration depth $\Delta^{Z,i}$ as the location at which the optical depth of a photon of frequency $\nu \gtrsim \nu_{\rm thr}^{Z,i}$ to absorption reaches unity, i.e.~
\be
1 = \int_{0}^{\Delta^{Z,i}}\rho_{\rm sh}\kappa_{\rm bf,\nu}^{Z,i}\left[1 +\rho_{\rm sh}\kappa_{\rm es}s \right]ds \approx \tau_{\rm abs}^{Z,i}(1 + \tau_{\rm es}^{Z,i}),
\label{eq:tau1}
\ee
where $s$ is the depth through the ionized layer, $\rho_{\rm sh}$ is the density of the shocked gas, and 
\be
\tau_{\rm abs}^{Z,i} \equiv \kappa_{\rm bf,\nu}^{Z,i}\rho_{\rm sh}\Delta^{Z,i}
\ee
is the optical depth through the layer to absorption, where 
\be
\kappa_{\rm bf,\nu}^{Z, i} \simeq \frac{X_A f_{\rm n}^{Z,i}}{A m_p}\sigma_{\rm bf,\nu}^{Z,i}
\label{eq:kappa}
\ee
is the absorption opacity and $\sigma_{\rm bf,\nu}^{Z,i}$ is the bound-free cross section.  The factor $1 + \tau_{\rm es}^{Z,i}$ in equation (\ref{eq:tau1}) accounts for the additional path-length traversed by the photon due to electron scattering,
where
\be
\tau_{\rm es}^{Z,i} = \rho_{\rm sh}\kappa_{\rm es}\Delta^{Z,i}
\ee
is the electron scattering optical depth through the ionized layer.  The neutral fraction $f_n^{Z,i}$ is determined by the balance between ionization and recombination
\be
f_n^{Z,i} = \left(1 + \frac{4\pi}{\alpha_{\rm rec}^{Z,i} n_e}\int\frac{J_{\nu}}{h\nu}\sigma^{Z,i}_{\rm bf}(\nu)d\nu\right)^{-1},
\label{eq:fn}
\ee
where $J_{\nu}$ is the mean intensity inside the nebula (eq.~[\ref{eq:Jnu}]); $n_e \approx \rho_{\rm sh}/2m_p$ is the electron density of the ejecta.  Here $\alpha_{\rm rec}^{Z,i}$ is the radiative recombination rate coefficient, which depends on the temperature of the electrons $T_e^{Z,i}$ in the ionized layer.  The electron temperature is set by Compton equilibrium $T_e^{Z,i} = T_{\rm C}$ (eq.~[\ref{eq:Tcompton}]), which also depends on the spectrum of photons present in the ionizing layer.  

The above discussion assumes that the penetration depth $\Delta^{Z,i}$ is identical for all photons with frequency $\nu \gtrsim \nu_{\rm thr}^{Z,i}$.  This is a reasonable approximation because most of the ionizing photons responsible for penetrating the neutral medium have $\nu \sim \nu_{\rm thr}^{Z,i}$ due to the decreasing number of nebular photons $\propto J_{\nu} \propto \nu^{-1}$ at higher frequencies.  The value of $\Delta^{Z,i}$ is also derived using the neutral fraction $f_{n}^{Z,i}$ (eq.~[\ref{eq:fn}]) calculated from the {\it unattenuated} nebular spectrum $J_{\nu}$.  A more detailed calculation, accounting for the attenuation of the ionizing radiation field with depth due to absorption (Appendix \ref{sec:hlike}), shows that photons with frequency $\nu \gtrsim 2\nu_{\rm thr}^{Z,i}$ in fact penetrate to a depth which is $\sim 2-3$ times greater than those with $\nu = \nu_{\rm thr}^{Z,i}$  (Fig.~\ref{fig:iondepth}).

As the ejecta expands and recombination becomes less efficient, $\Delta^{Z,i}$ increases with time.  Complete ionization of the ejecta (`ionization break-out') is achieved once $\Delta^{Z,i}$ exceeds the ejecta thickness $\Delta_{\rm sh}$.  Because $\Delta^{Z,i}/\Delta_{\rm sh}$ is greatest for atomic species with low ionization energies, break-out (when it occurs) begins first at low frequencies and then moves to higher frequencies with time.  This transition is not achieved continuously, but instead occurs as discrete frequency windows are opened by the sequential ionization of species with increasingly higher ionization energies.


Finally, considering harder X-rays, inelastic electron scattering suppresses radiation above frequency
\be
h \nu \sim \frac{m_{\rm e}c^{2}}{(\tau_{\rm es}^{\rm ej})^{2}} \sim 0.5v_9^{2}\left(\frac{t}{t_{\rm d,0}}\right)^{4} {\rm keV},
\label{eq:hnuies}
\ee
where $(\tau_{\rm es}^{\rm ej})^{2}$ is the number of scatterings experienced by a photon and
\be
\tau^{\rm ej}_{\rm es} \simeq \frac{3-\delta}{4\pi}\frac{M_{\rm ej}\kappa_{\rm es}}{R_{\rm ej}^{2}} \approx 30v_{9}^{-1}\left(\frac{t}{t_{\rm d,0}}\right)^{-2}
\label{eq:tau_ej}
\ee
is the characteristic optical depth of the ejecta.  The first equality in equation (\ref{eq:tau_ej}) is valid only for $\delta \lesssim 1$ and the numerical estimate assumes $\delta = 0$, consistent with our definition of $t_{\rm d,0}$ (eq.~[\ref{eq:td}]).  Inelastic scattering thus prevents hard $\gg 10$ keV X-rays from escaping until times $t \gtrsim $ few $t_{\rm d,0}$.      

The interaction between the radiation of the nebula and the ejecta is usefully characterized by the `ionization parameter' (e.g.~\citealt{Ross+99})
\be
\xi \equiv \frac{4\pi F}{n_{\rm e}} \approx \frac{2\pi c E_{\rm nth}m_p}{M_{\rm ej}}\simeq 10^{4}L_{45}M_{3}^{-1/2}v_{9}^{1/2}\left(\frac{t}{t_{\rm d,0}}\right){\rm \,\,erg \,\,cm \,\,s^{-1}},
\label{eq:ion_param}
\ee
where $F \simeq cE_{\rm nth}/4 V_{\rm n}$ is the total flux of ionizing photons; we approximate the electron density as $n_{\rm e} \approx M_{\rm ej}/2m_p V_{\rm ej}$; and $E_{\rm nth}$ is again estimated using equation (\ref{eq:Ent_eq}) with $\bar{\mathcal{A}}_{\nu} \sim 0.5$.  The characteristic values $\xi \sim 10^{3}-10^{4}$ in young millisecond PWNe are similar to those achieved in corona-irradiated black hole accretion disks, for which $\xi = 7\times 10^{4}f_{\rm edd}^{3}$, where $f_{\rm edd}$ is the accretion Eddington fraction (\citealt{Ross&Fabian93}).  

\section{Evolutionary Model}
\label{sec:toy}

Having introduced the key physical ingredients, we now describe a simple one-zone model for the evolution of the nebula and the ionization structure of the ejecta, which we use to quantify the conditions for ionization break-out in $\S\ref{sec:results}$.

\subsection{Size of Nebula and Shocked Ejecta}
\label{sec:size}

The pressure $P_{\rm n} = E/3V_{\rm n}$ of the nebula drives a strong shock into the expanding supernova ejecta with a characteristic velocity, as measured in the frame co-moving with the SN ejecta, given by
\be v_{\rm sh}(t) = \sqrt{\frac{7}{6}\frac{P_{\rm n}}{\rho_{\rm ej}(R_{\rm n}, t)}},
\label{eq:vsh}
\ee 
where $\rho_{\rm ej}(R_{\rm n},t)$ is the unperturbed density profile of the ejecta  (eq.~[\ref{eq:rho0}]) .

The shocked SN ejecta is swept up into a thin shell, which is separated by a contact discontinuity from the hot, low-density interior of the nebula and which contains a fraction
\be
f_{\rm sh} = \left(\frac{R_{\rm n}}{R_{\rm ej}}\right)^{3-\delta}
\label{eq:fsh}
\ee
of the total ejecta mass (Fig.~\ref{fig:compare}).  

The thickness of the thin shell $\Delta_{\rm sh}$ is determined by equating the mass of the shell $M_{\rm sh} = f_{\rm sh}M_{\rm ej}$ to the product of its volume $4\pi R_n^{2}\Delta_{\rm sh}$ and its density $\rho_{\rm sh} \simeq 7 \rho_{\rm ej}(R_n,t)$, where the factor of $7$ is the compression ratio for a strong $\gamma = 4/3$ shock.  This gives
\be
\frac{\Delta_{\rm sh}}{R_n} = \frac{1}{(3-\delta)7}
\label{eq:delta_sh}
\ee
The column density of the shocked ejecta shell is given by
\be
\Sigma_{\rm sh} = \frac{f_{\rm sh}M_{\rm ej}}{4\pi R_{\rm n}^{2}},
\label{eq:sigmash}
\ee
with a corresponding electron scattering optical depth
\be
\tau^{\rm sh}_{\rm es} = \Sigma_{\rm sh}\kappa_{\rm es}.
\label{eq:tauessh}
\ee

The radius of the nebula and shocked ejecta grows according to
\be
\frac{dR_{\rm n}}{dt} = v_{\rm sh}(t) + \frac{R_{\rm n}}{t},
\label{eq:dRndt}
\ee
where the last term is the velocity of the ejecta at the position which is just about to be swept up by the shock (this accounts for the fact that $v_{\rm sh}$ is measured co-moving with the ejecta).  

If the pulsar wind is sufficiently luminous, then the nebular radius will overtake the outer radius of the ejecta ($R_{\rm n} \gtrsim R_{\rm ej}$).  In this case the entire ejecta is swept into a thin shell of velocity 
\be
v_{\rm ej}^{\rm f} \simeq \left(\frac{2\int_0^{t} L_{\rm sd}dt'}{M_{\rm ej}} + v_{\rm ej}^{2}\right)^{1/2},
\label{eq:vejf}
\ee where $\int_0^{t}L_{\rm sd}dt'$ is the rotational energy injected by the pulsar up to that point.  We account for this case in our calculations by taking
\be
\frac{dR_{\rm ej}}{dt} = \frac{dR_{\rm n}}{dt} = v_{\rm ej}^{\rm f}\,\,\,\,\,\, {\rm if}\,\,R_{\rm n} \ge R_{\rm cr},
\label{eq:dRejdt}
\ee
where $R_{\rm cr}$ is the `crossing radius', which is defined as the radius at which $R_{\rm n} = R_{\rm ej}$ according to the evolution predicted by equation (\ref{eq:dRndt}).  We thus take $R_{\rm ej} \simeq R_{\rm n}$ for $R_{\rm n}, R_{\rm ej} > R_{\rm cr}$.
 
\subsection{Total Radiation Energy}
\label{sec:energy}

The total energy of the system in radiation (thermal and non-thermal) evolves according to
\begin{eqnarray}
\frac{dE}{dt} = -\frac{E}{R_{\rm ej}}\frac{dR_{\rm ej}}{dt} + L_{\rm sd}(t) + \frac{4\pi R_{\rm n}^{2}v_{\rm sh}}{7}\frac{E}{V_{\rm n}} -  L_{\rm rad}(t),
\label{eq:dEdt}
\end{eqnarray}
The first term accounts for losses due to adiabatic expansion in a radiation-dominated gas.  The second term accounts for energy input from the pulsar, neglecting the small fraction of the energy deposited into the magnetic field or into the rest mass of pairs (eq.~[\ref{eq:taueseq}]).  As a reminder, we assume that the initial Poynting flux of the pulsar wind is efficiently dissipated within the nebula by shocks or magnetic reconnection (eq.~[\ref{eq:Bn}]).  The third term accounts for thermal energy deposited by the dissipation of kinetic energy at the shock (eq.~[\ref{eq:vsh}]).  The final term
\be
L_{\rm rad} = \frac{E_{\rm th}}{t_{\rm d}} \equiv L_{\rm SN},
\label{eq:L_SN}
\ee
accounts for radiative losses from the system, as is responsible for powering the supernova luminosity $L_{\rm SN}$, where
\be
t_{\rm d} =  \left(\tau_{\rm es}^{\rm ej} + 1\right)\frac{R_{\rm ej}}{c}
\label{eq:td2}
\ee
is the characteristic photon diffusion timescale through the ejecta and $\tau^{\rm ej}_{\rm es}$ is the total optical depth of the ejecta (eq.~[\ref{eq:tau_ej}]).  The factor of unity inside the parentheses extrapolates the cooling timescale smoothly from the optically-thick to optically-thin cases; in the latter case $t_{\rm d}$ becomes the light crossing time.  As discussed in $\S\ref{sec:opacity}$, the optical depth is calculated assuming electron scattering opacity since this is reasonably accurate at optical wavelengths, regardless of whether the ejecta is in fact fully ionized, or whether it is neutral and Doppler-broadened lines dominate the opacity. 

Note that only the thermal bath $E_{\rm th}$ contributes to radiative losses from the nebula in equation (\ref{eq:L_SN}).  This is justified at early times because the nebula is highly opaque to UV/X-ray photons.  At later times, however, progressively higher ionization states in the ejecta become fully ionized, allowing energy loss to also occur directly at UV/X-ray wavelengths.  Equation (\ref{eq:L_SN}) is nevertheless a reasonable approximation for our purposes since (1) the energy density of the nebula at UV-X-ray frequencies is at most comparable to that at optical frequencies, so neglecting this additional energy loss results in a modest loss of accuracy in calculating the nebula dynamics; (2) we are predominantly interested in following the nebular evolution up until the point of X-ray transparency, in order to determine the requisite conditions for such break-out to occur.

If the pulsar spin-down time $t_{\rm sd}$ is somewhat less than, or comparable to, the characteristic diffusion timescale $t_{\rm d,0}$, then $L_{\rm SN}(t)$ peaks on the characteristic timescale $\sim t_{\rm d,0} \sim$ month (eq.~[\ref{eq:td}]), although the duration of the peak can last significantly longer if $t_{\rm sd} \gg t_{\rm d,0}$.

\subsection{Nebula Spectrum}
\label{sec:nebspectra}

The spectrum of non-thermal radiation in the nebula evolves according to
\be
\frac{dE_{\rm nth,\nu}}{dt} =  -\frac{E_{\rm nth,\nu}}{R_{\rm ej}}\frac{dR_{\rm ej}}{dt} + \dot{E}_{\rm sd,\nu} - (1-\mathcal{A}_{\nu})\frac{E_{\rm nth,\nu}}{t_{\rm d}^{\rm n}} ,
\label{eq:photon_evo}
\ee
while the thermal bath of the nebula and ejecta evolves as
\be
\frac{dE_{\rm th}}{dt} = -\frac{E_{\rm th}}{R_{\rm ej}}\frac{dR_{\rm ej}}{dt}  - L_{\rm rad} + \frac{4\pi R_{\rm n}^{2}v_{\rm sh}}{7}\frac{E}{V_{\rm n}} + \int (1-\mathcal{A}_{\nu})\frac{E_{\rm nth,\nu}}{t_{\rm d}^{\rm n}} d\nu.
\label{eq:thermo_evo}
\ee
The first term in both equations again accounts for PdV losses assuming $R_{\rm ej} \approx R_{\rm n}$.\footnote{The first term in equation (\ref{eq:photon_evo}) does not properly take into account the {\it redistribution} of photon energy due to adiabatic expansion.  Nevertheless, this expression is valid for a flat spectrum with $\nu E_{\rm nth, \nu} \sim$ constant, as approximately characterizes the nebular spectrum in our calculations.}  The second term in equation (\ref{eq:photon_evo})   
\be
\dot{E}_{\rm sd,\nu} \simeq \frac{L_{\rm sd}}{14\nu}\,\,\,\,\,\,\,\,(3kT_{\rm th}  \sim 1 \,\,{\rm eV} \lesssim h\nu \lesssim 1 \,\,{\rm MeV})
\label{eq:edotsd}
\ee
accounts for the photon spectrum injected by the pair cascade ($\S\ref{sec:cascade}$), normalized such that $\int \dot{E}_{\rm sd,\nu}d\nu$ equals the total pulsar power $L_{\rm sd}(t)$ (eq.~[\ref{eq:Lsd}]).  The second and third term in equation (\ref{eq:thermo_evo}) accounts for radiative losses and shock heating, as was already introduced following equation (\ref{eq:dEdt}).

The last terms in equations (\ref{eq:photon_evo}) and (\ref{eq:thermo_evo}) account for the absorption of photons by the ionized inner layer of the ejecta, where 
\be t_{\rm d}^{\rm n} \simeq \frac{R_{\rm n}}{c}(\tau_{\rm es}^{\rm n} + 1)
\label{eq:tdiffn}
\ee
is the timescale for photons to propagate through the nebula and $\tau_{\rm es}^{\rm n}$ is the pair optical depth through the nebula (eq.~[\ref{eq:taueseq}]).  

The factor $1-\mathcal{A}_{\nu}$ is the probability that a photon is absorbed, or loses significant energy to down-scattering, in the ionized layer instead of being reflected back into the nebula, where $\mathcal{A}_{\nu}$ is the albedo at frequency $\nu$.   Energy absorbed by the ejecta is assumed to be completely thermalized.  The albedo is a function of the [frequency-dependent] ratio of the absorption and scattering opacity 
\be \eta \equiv \frac{\kappa_{\rm abs,\nu}}{\kappa_{\rm es}},
\label{eq:eta1}
\ee 
in the ionized layer dominated by the atomic species (Z,i) of interest ($\S\ref{sec:opacity}$),
where 
\be 
\kappa_{\rm abs,\nu} = \kappa_{\rm bf,\nu}^{Z,i} + \kappa_{\rm ie},
\ee
$\kappa_{\rm bf,\nu}^{Z,i}$ is the bound-free opacity (eq.~[\ref{eq:kappa}]) of the dominant absorbing species ($Z,i$), and
\be
\kappa_{\rm ie} = \kappa_{\rm es}\frac{h\nu}{m_{\rm e}c^{2}}
\ee
is the absorptive opacity for a photon of frequency $\nu$ due to inelastic down-scattering.  This expression follows from the fact that a photon loses a fraction $\sim h\nu/m_e c^{2}$ of its energy per scattering.

\begin{figure}
\includegraphics[width=0.5\textwidth]{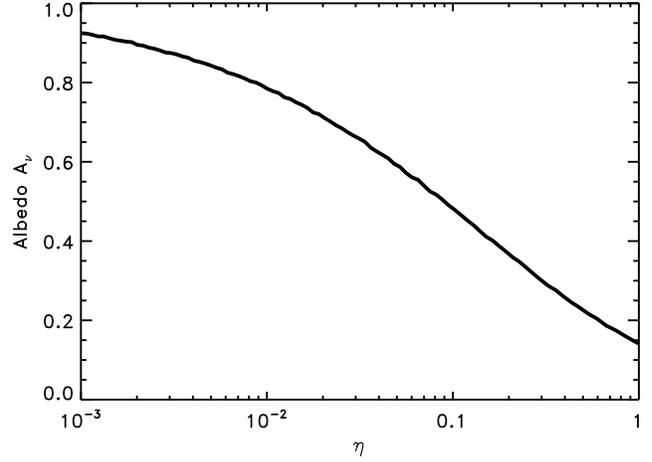}
\caption{Albedo of the ionized ejecta as a function of $\eta \equiv \kappa_{\rm abs,\nu}/\kappa_{\rm es}$ (eq.~[\ref{eq:eta1}]), the ratio of the absorptive and electron scattering opacity in the ionized layer (see text).} 
\label{fig:albedo}
\end{figure}

We calculate $\mathcal{A}_{\nu}(\eta)$ using a Monte Carlo procedure by which the fate of a large number of photons injected into one side of the slab is followed, counting the fraction absorbed, reflected, or transmitted through the slab.  For a given value of $\eta$, the thickness of the ionized layer used in this calculation is determined by the condition that the effective optical depth of the layer equal unity (eq.~[\ref{eq:tau1}]).  The results are shown in Figure \ref{fig:albedo}.

Our simple prescription for directly depositing the photon energy absorbed by the ionized layer into the thermal bath does not take into account details such as line re-emission, as would be captured by a complete photo-ionization model.  Nevertheless, photo-electrons do carry a sizable fraction of the absorbed radiation, and these do thermalize.  Our method should thus capture the continuum flux evolution reasonably accurately, which is of greatest interest here because it controls the ionization state of the ejecta.   

\subsection{Ionization Fronts}

As the ejecta expands and its density decreases, multiple ionization fronts propagate outwards through the shocked ejecta.  Because the radiative recombination rate increases with density, and since the shocked inner ejecta shell has a higher density than the unshocked outer ejecta (by a factor $\gtrsim 7$; $\S\ref{sec:size}$), the shocked ejecta will be more neutral and hence will dominate the total optical depth.  As a result, if the ionization front reaches the front of the shock before the shock reaches the ejecta surface ($R_{\rm n} < R_{\rm ej}$), then subsequent ionization of the outer (unshocked) ejecta occurs rapidly in comparison.  

Using the definition of the penetration depth of the ionization front $\Delta^{Z,i}$ as the location of an effective optical depth of unity (eq.~[\ref{eq:tau1}]), one can write (cf.~Appendix \ref{sec:hlike})
\be
\frac{\Delta^{Z,i}}{\Delta_{\rm sh}} \approx 2\left(\frac{\sqrt{1 + 4 \eta_{\rm thr}^{-1}} - 1}{2 \tau_{\rm es}^{\rm sh}}\right),
\label{eq:ionthick}
\ee
where $\Delta_{\rm sh}$ is the total width of the shocked ejecta; $\tau_{\rm es}^{\rm sh}$ (eq.~[\ref{eq:tauessh}]) is the electron scattering optical depth through the shocked shell; and
\be
\eta_{\rm thr} \equiv \left.\frac{{\kappa}_{\rm abs,\nu}^{Z,i}}{\kappa_{\rm es}}\right|_{\nu =  \nu_{\rm thr}^{Z,i}}
\label{eq:eta}
\ee
is the ratio of absorptive and scattering opacities (eq.~[\ref{eq:eta1}]) at the ionization threshold frequency $\nu_{\rm thr}^{Z,i}$.  The prefactor of 2 in equation (\ref{eq:ionthick}) accounts for the fact that photons with $\nu \gtrsim 2 \nu_{\rm thr}$ penetrate to a depth that is $\sim 2$ times larger than photons at the ionization threshold frequency (Figure \ref{fig:iondepth}).  We define the penetration depth as that of photons with frequency $\nu \gtrsim 2\nu_{\rm thr}$ since this represents a typical location within the frequency band of relevance and we are ultimately interested in the conditions required for the bulk of the nebular luminosity to escape.  

Ionization break-out occurs when the right hand side of equation (\ref{eq:ionthick}) exceeds unity.  After this point, most photons in this band may freely escape, other than the minimal delay due to electron scattering.  In cases when the shock reaches the front of the ejecta ($R_{\rm n} \rightarrow R_{\rm ej}$) the shocked layer thickness is no longer $\Delta_{\rm sh} \simeq R_{\rm n}/7(3-\delta)$.  Nevertheless, the right hand side of equation (\ref{eq:ionthick}) reaching unity remains a valid condition for break-out in this case as well.  

The absorptive opacity in equation (\ref{eq:eta}) is calculated from equations (\ref{eq:kappa}) and (\ref{eq:fn}) for $\nu = \nu_{\rm thr}^{Z,i}$.  The neutral fraction $f_{\rm n}^{Z,i}$ is calculated from equation (\ref{eq:fn}) using the electron density of the shocked ejecta $n_e \simeq \rho_{\rm sh}/2m_p$.  The density of gas in the ionized layer is given by
\begin{eqnarray}
\rho_{\rm sh} &=& 7 \rho_0(R_{\rm n},t)\,\,\,\,\,\,\,\,\, (R_{\rm n} \lesssim R_{\rm ej}) \nonumber \\
	& = & \frac{M_{\rm ej}}{V_{\rm ej}}\left(\frac{18 - 7\delta}{3}\exp\left[-\frac{(R_{\rm ej}-R_{\rm cr})}{R_{\rm cr}}\right] + 1\right)\,\,\,\,\,\,\,\,\,(R_{\rm n} = R_{\rm ej})
\label{eq:rhosh}
\end{eqnarray}
The second line accounts for cases in which the shock reaches the ejecta surface (at radius $R_{\rm cr}$; see eq.~[\ref{eq:dRejdt}]) by smoothly interpolating between the initially higher density of the thin shocked layer (when $R_{\rm ej} \sim R_{\rm ej}$) and the eventual homologous expansion at later times ($R_{\rm ej} \gg R_{\rm cr}$).  

The large number of elements and ionization states comprising the ejecta make calculating the time-dependent structure of the ionized layers a significant undertaking.  A similar problem arises in X-ray irradiation of accretion disks (Fig.~\ref{fig:compare}), which has motivated an extensive literature of detailed calculations of X-ray reprocessing (e.g.~\citealt{Ross&Fabian93}; \citealt{Zycki+94}; \citealt{Stern+95}; \citealt{Ross+99}; \citealt{Ross&Fabian05}; \citealt{Kallman+04}).  We do not strive for such detail here, but instead estimate the ionized layer thickness using a limited set of species that appear to dominate the growth of the ionization front in the frequency ranges of relevance.  

We focus on the ionization states of oxygen due to its large abundance in the ejecta ($X_{\rm O} = 0.59$ in our fiducial model; $\S\ref{sec:ejecta}$).  To cover energies above the ionization threshold of hydrogen-like oxygen ($h \nu \gg $ keV), we also include Fe ions since iron is relatively abundant ($X_{\rm Fe} = 0.05$), difficult to ionize (Appendix $\ref{sec:hlike}$), and has ionization energies up to $\sim 10$ keV.  Recombination coefficients $\alpha_{\rm rec}(T_{\rm C}^{Z,i})$ are taken from \citet{Nahar99} for oxygen and from \citet{Woods+82} for iron.  Analytic fits to the bound-free cross sections $\sigma_{\rm bf,\nu}^{Z,i}$ are taken from \citet{Verner+96}. 

The recombination coefficients are calculated at the Compton temperature $T_{\rm C}^{\rm Z,i}$ of electrons in the ionization layer.  Note that $T_{\rm C}^{Z,i}$ may be significantly lower than the Compton temperature in the nebula.  It is calculated using equation (\ref{eq:Tcompton}), but performing the integration over just the frequency window(s) corresponding to photons present in the ionizing layer of interest.  At a minimum this includes frequencies from the thermal bath up to a maximum frequency set by the ionization threshold of the next ionization state.  It also potentially includes much harder photons, extending from the minimum energy $\sim 30$ keV not suppressed by bound-free absorption (by iron), up to a maximum photon energy
\be
h\nu_{\rm max}^{\rm C} = \frac{m_e c^{2}}{(\tau_{\rm es}^{Z,i})^{2}} = \frac{4 m_e c^{2}}{\left(\sqrt{1 + 4\eta_{\rm thr}^{-1}}-1\right)^{2}}
\label{eq:numax}
\ee
above which photons are suppressed by inelastic electron scattering (eq.~[\ref{eq:hnuies}]), where we have used the definition of the ionized layer from equations (\ref{eq:tau2}) and (\ref{eq:dth}).  Note that this renders the expression for $\eta_{\rm thr} \propto \alpha_{\rm rec}(T_{\rm C}^{Z,i}[\eta_{\rm thr}])$ implicit.  

\begin{figure}
\includegraphics[width=0.5\textwidth]{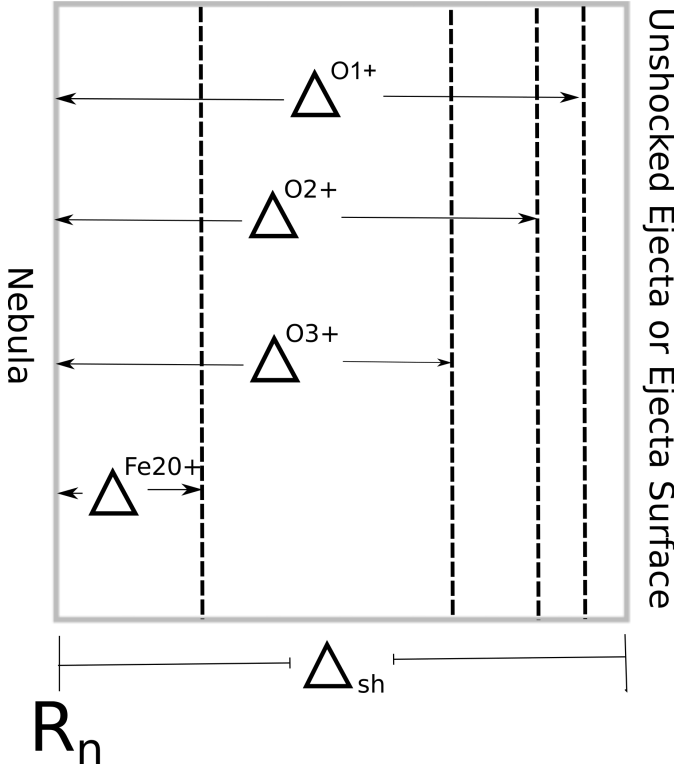}
\caption{Schematic drawing of the ionization structure of the shocked layer separating the nebula on the left from the unshocked ejecta, or the surface of the ejecta (after the shock has reached the surface), on the right.  Each atomic state $(Z,i)$ is ionized to a depth $\Delta^{Z,i}$ through the shocked layer of total width $\Delta_{\rm sh}$, as determined by equation (\ref{eq:ionthick}).  Lower ionization states generally penetrate further (higher $\Delta^{Z,i}$) due in part to the greater number of ionizing photons in the nebula at low energies (Appendix \ref{sec:hlike}).  Once the shocked layer is completely ionized ($\Delta^{Z,i} \rightarrow \Delta_{\rm sh}$), photons with frequencies above the threshold ionization energy of that ionization state are free to diffuse out of the ejecta unattenuated (`ionization break-out').} 
\label{fig:layers}
\end{figure}

Since some species are more easily ionized than others, only a limited set of ionization states are relevant in setting the ionization depth across a given range of photon energies.  These most important ions are determined by first calculating $\Delta^{Z,i}$ for {\it all} ionization states of oxygen and iron, assuming that photons at $\nu \sim \nu_{\rm thr}^{Z,i}$ propagate through the ejecta unattenuated, except due to absorption by species $(Z,i$).  Then, starting at low frequencies and moving to progressively higher frequencies, one `eliminates' those species with a larger value of $\Delta^{Z,i}$ than the penetration depth allowed by the previous element with the lower ionization frequency.  Applying this procedure results in the following set of {\it relevant} ionization states: O$^{1+}$ (14$-$35 eV), O$^{2+}$ (35$-$55 eV), O$^{3+}$ (55 eV$-$1.58 keV), Fe$^{+20}$ ($>$ 1.58 keV), where the relevant photon energy range is given in parentheses.  Figure \ref{fig:layers} is a schematic illustration of the ionization structure of the shocked ejecta.    

We acknowledge that this simple-minded procedure is ultimately no substitute for a time-dependent photo-ionization calculation.  It nevertheless provides a basic estimate of what conditions are required for UV/X-ray photons of a given energy to directly escape the nebula.  Also note that our calculation does not include corrections for the finite propagation time of ionizing photons; we instead assume that ionization balance is achieved instantaneously as the conditions in the nebula change.  This is a good approximation at times $t \gtrsim t_{\rm d,0}$ (eq.~[\ref{eq:td}]), as is usually satisfied for UV [and always satisfied for X-ray] ionization break-out (Fig.~\ref{fig:contour}, \ref{fig:contour2}).

\subsection{Computational Procedure}
\label{sec:comp}

\begin{figure}
\centering
\subfigure{
\includegraphics[width = 0.48\textwidth]{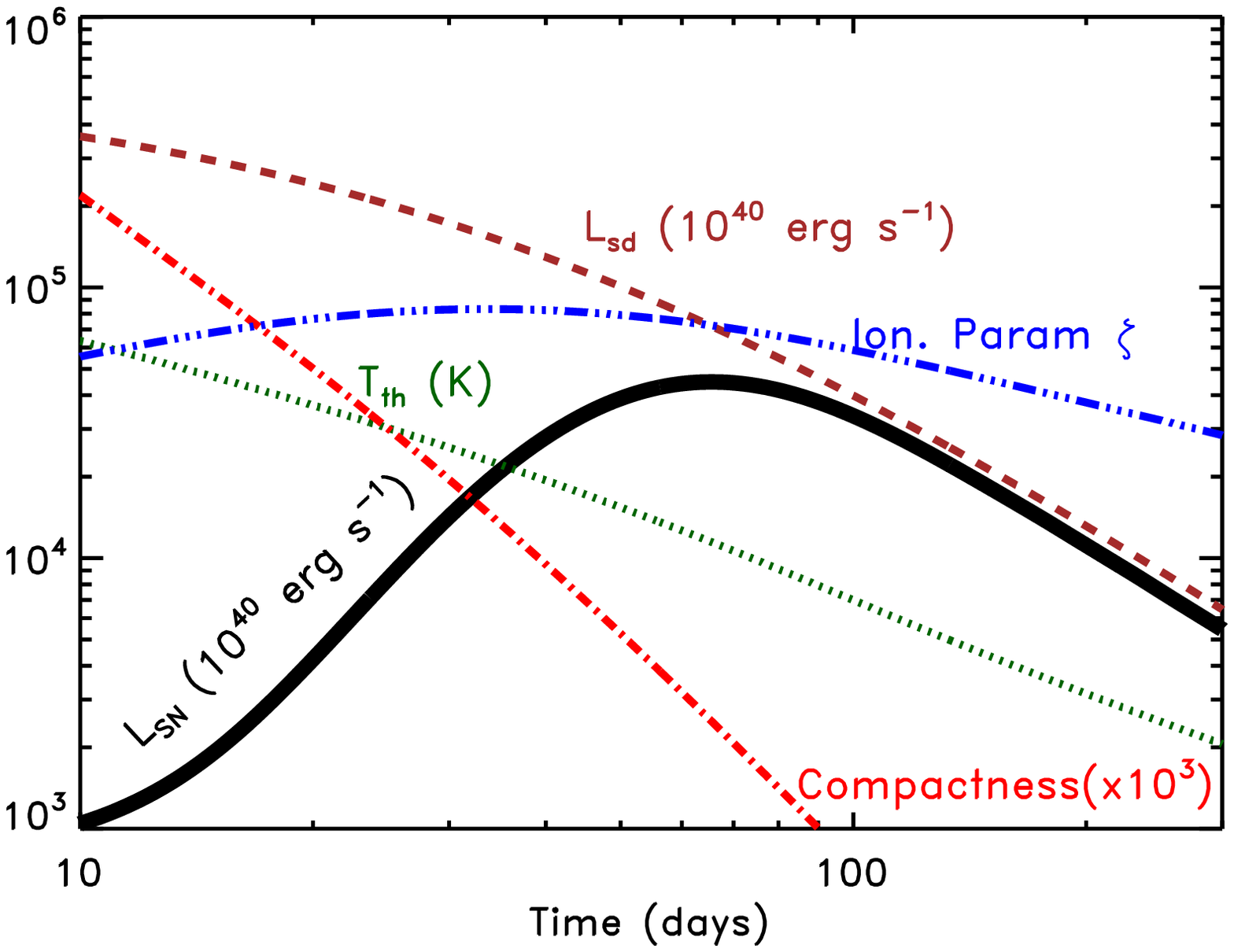}}
\subfigure{
\includegraphics[width = 0.48\textwidth]{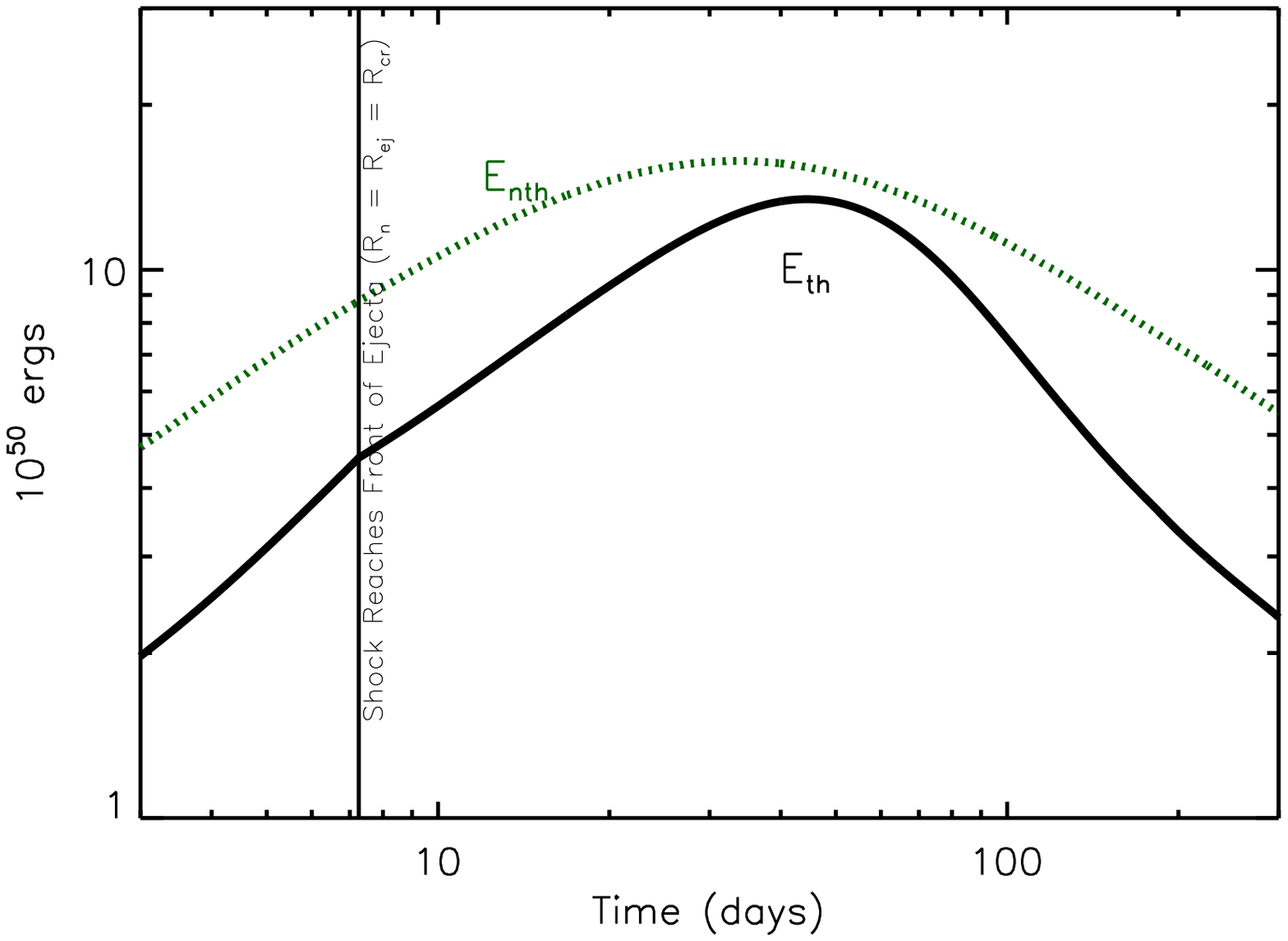}}
\caption[]{{\it \normalsize Top Panel:} Various quantities as a function of time since the supernova explosion, calculated for a pulsar with dipole field $B_{\rm d} =  10^{13}$ G and initial rotation period $P = 1$ ms; ejecta mass $M_{\rm ej} = 3M_{\odot}$, initial velocity $v_{\rm ej} = 10^{9}$ cm s$^{-1}$, and density profile $\delta = 1$.  Quantities shown include the spin-down luminosity of the pulsar $L_{\rm sd}$ ({\it dashed brown}; eq.~[\ref{eq:Lsd}]); the bolometric luminosity of the thermal supernova emission $L_{\rm SN}$ ({\it solid black}; eq.~[\ref{eq:L_SN}]); temperature of thermal bath in nebula $T_{\rm th}$ ({\it dotted green}; eq.~[\ref{eq:Tn}]); compactness of nebula $\ell$ ({\it red dot-dashed}; eq.~[\ref{eq:compactness}]); and ionization parameter of interaction between nebular flux and ejecta walls in units of erg cm$^{-1}$ s$^{-1}$ ({\it blue triple dot-dashed}; eq.~[\ref{eq:ion_param}]).  {\it \normalsize Bottom Panel:} Contributions to the total energy budget, for the same calculation shown in above panel, including the thermal energy of the ejecta and nebula $E_{\rm th}$ ({\it black solid}) and the non-thermal energy of nebula $E_{\rm nth}$ ({\it green dashed}).}
\label{fig:fig1}
\end{figure}

We now briefly summarize how the ingredients described above are assembled to give a numerical model for the evolution of the nebula/ejecta and ionization fronts.  The radius of the nebula and the ejecta are evolved according to equations
(\ref{eq:dRndt}) and (\ref{eq:dRejdt}), as driven by the nebula pressure and coupled by their shock interaction, starting the nebula from a compact initial state $R_{\rm n} \ll R_{\rm ej}$ at $t = 0$.  The total radiation energy of the nebula and ejecta are evolved according to equation (\ref{eq:dEdt}).  The thermal bath and non-thermal radiation field are evolved separately by equations (\ref{eq:thermo_evo}) and (\ref{eq:photon_evo}), as driven by energy injection from the pulsar and coupled by the frequency-dependent albedo of the walls.  The albedo in turn depends on the ratio of absorptive and scattering opacities $\eta$ (eq.~[\ref{eq:eta1}]) in the ionized layer at the frequency of interest (Fig.~\ref{fig:albedo}), which is calculated based on the bound-free opacity of the nearest ion ($Z,i$) of relevance according to equation (eq.~[\ref{eq:eta1}]).  The condition for break-out (eq.~[\ref{eq:ionthick}]) in a given frequency band is determined by $\eta_{\rm thr}$ (eq.~[\ref{eq:eta}]) at the threshold frequency of the ion ($Z,i$) corresponding to that band.  Again note that $\eta_{\rm thr}$ is solved via an implicit procedure since the recombination rate depends on the (Compton) temperature of electrons, which itself depends on the Thomson depth of the ionizing layer via the maximum energy of photons allowed to penetrate the layer despite losses to inelastic scattering (eq.~[\ref{eq:numax}]).    

The problem is completely specified by the properties of the pulsar $(P_{-3}, B_{\rm 13})$ and the ejecta $(v_9, M_3, \delta, X_A)$.

\section{Results}
\label{sec:results}

\begin{figure}
\centering
\includegraphics[width = 0.48\textwidth]{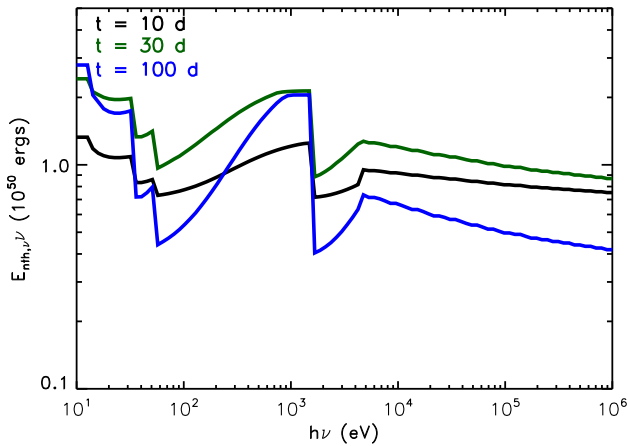}
\caption[]{
Energy distribution $\nu E_{\rm nth,\nu}$ of non-thermal radiation inside nebula for the same calculation in Figure \ref{fig:fig1}, shown at various times as marked.  }
\label{fig:spectra}
\end{figure}

\begin{figure*}
\centering
\includegraphics[width = 0.6\textwidth]{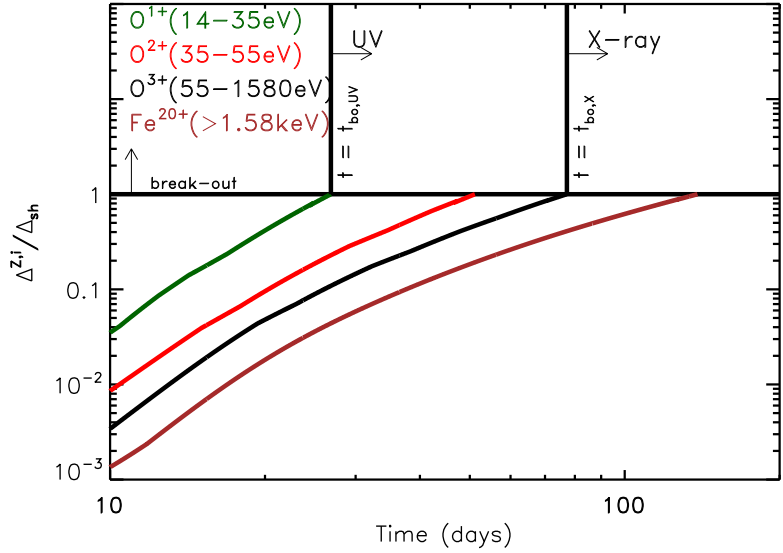}
\caption[]{Thickness of the ionized layer $\Delta^{Z,i}$ relative to the width of the shocked region $\Delta_{\rm sh}$ (eq.~[\ref{eq:ionthick}]) for those ions that dominate the ionization front in a given photon energy range, for the same calculation shown in Figures \ref{fig:fig1} and \ref{fig:spectra}.  Ions of oxygen ({\it solid}) and iron ({\it dashed}) are denoted by different colors as indicated.  Ionization break-out ($\Delta^{Z,i}/\Delta_{\rm sh} = 1$) occurs first at UV energies and later at soft X-ray energies. }
\label{fig:ionthick1}
\end{figure*}

Figures \ref{fig:fig1}$-$\ref{fig:ionthick1} show our results calculated for a pulsar with magnetic field $B_{\rm d} = 10^{13}$ G and initial spin period $P = 1$ ms (spin-down time $t_{\rm sd} \sim 1$ month), and for fiducial properties of the supernova ejecta ($M_{3} = v_9 = \delta = 1$; characteristic diffusion timescale $t_{\rm d,0} \sim$ 1 month).  The top panel of Figure \ref{fig:fig1} shows the time evolution of various quantities, including the nebula compactness $\ell$ (eq.~[\ref{eq:compactness}]); the ionization parameter of the interaction between the nebular X-ray field and the ejecta walls $\xi$ (eq.~[\ref{eq:ion_param}]); and the bolometric thermal luminosity of the supernova $L_{\rm SN}$ (eq.~[\ref{eq:L_SN}]).  The compactness decreases monotonically with time, but remains sufficiently high $\ell \gtrsim 1$ to sustain a pair cascade inside the nebula at times $t \lesssim$ 80 days.  The ionization parameter $\xi \sim 10^{4}-10^{5}$ (eq.~[\ref{eq:ion_param}]) is similar to that characterizing the interaction between the disk and corona of an Eddington-accreting black hole (Fig.~\ref{fig:compare}; e.g.~\citealt{Ross+99}).    

The supernova luminosity $L_{\rm SN}$ peaks at $\sim 5\times 10^{44}$ erg s$^{-1}$ on a timescale of $\sim 60$ days after core collapse, with a total radiated energy $E_{\rm rad} = \int L_{\rm SN}dt \simeq 7\times 10^{51}$ ergs.  Such prodigious radiative output is the hallmark of a SLSN.  The bottom panel of Figure \ref{fig:fig1} shows the evolution of the energy budget of the system, including the total thermal energy of the system $E_{\rm th}$ and the total non-thermal energy of the nebula $E_{\rm nth} \equiv \int E_{\rm nth,\nu}d\nu$ (the UV/X-ray tail).  The thermal energy initially rises rapidly due to the kinetic energy dissipated by the shock crossing the ejecta.  The shock reaches the surface of the ejecta ($R_{\rm n} = R_{\rm ej} = R_{\rm cr}$) at $t \sim 7$ days, after which time all additional radiation originates from the pulsar.   Of the non-thermal radiation injected into the nebula by the pulsar ($\S\ref{sec:cascade}$), a sizable fraction ($\sim 1/3$) is deposited into the thermal bath via the absorption and re-emission of hard radiation by the ejecta walls.  Thermal energy subsequently diffuses out of the ejecta, powering the bulk of the supernova emission.  The supernova light curve peaks on a characteristic timescale $\gtrsim t_{\rm d,0} \sim t_{\rm sd}$, which for this example is comparable to the pulsar spin-down time.

The nebula maintains significant energy in non-thermal radiation until late times.  Figure \ref{fig:spectra} shows snapshots in the evolution of the spectrum of non-thermal radiation $\nu E_{\rm nth,\nu}$ at several times, revealing the complex spectrum imprinted by selective absorption due to the frequency-dependent albedo of the ejecta walls (cf.~Fig.~\ref{fig:albedo} and eq.~[\ref{eq:eta1}]).  Like the thermal bath, this non-thermal energy is free to leak out once the ejecta becomes fully ionized and transparent to higher frequency photons.  

Figure \ref{fig:ionthick1} shows the thickness of the ionization layer $\Delta^{Z,i}/\Delta_{\rm sh}$ for elements and ionization states of interest as a function of time, for the same calculation shown in Figures \ref{fig:fig1} and \ref{fig:spectra}.  For each species $\Delta^{Z,i}$ increases monotonically with time, as the electron density (and hence recombination rate) decreases.   The O$^{1+}$ layer reaches the surface of the ejecta first ($\Delta^{Z,i} \gtrsim \Delta_{\rm sh}$) first, within $\sim 25$ days of the explosion.  Photons with $h\nu \lesssim 35$ eV can thus escape the nebula on the diffusion timescale over which the bulk of the optical photons are released.  In $\S\ref{sec:UV}$ we speculate on the resulting implications for the UV colors of SLSNe-I.   

The fronts of higher ionization states of oxygen (O$^{3+}$ and beyond) reach the ejecta surface at $t \gtrsim 80$ days, after which time X-rays with energies $h \nu  \gtrsim 0.1-1$ keV can also escape the nebula.  The energy previously trapped in non-thermal radiation at the time of break-out, $E_{\rm nth} \gtrsim 10^{51}$ ergs, is quite significant.  Finally, fronts of high ionization states of iron (Fe$^{20+}$) do not reach the surface until significantly later, delaying the escape of hard X-rays.  Although the abundance of iron in the ejecta is an order of magnitude lower than that of oxygen, the fewer ionizing photons at high energies generally makes ionizing elements with higher nuclear charge $Z$ more difficult (Appendix \ref{sec:hlike}).

\begin{figure}
\centering
\subfigure{
\includegraphics[width = 0.48\textwidth]{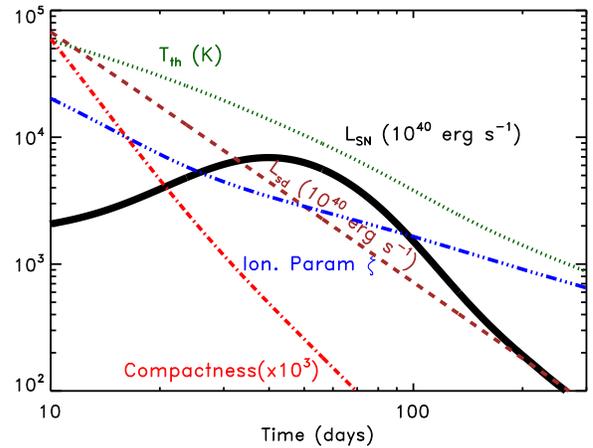}}
\subfigure{
\includegraphics[width = 0.48\textwidth]{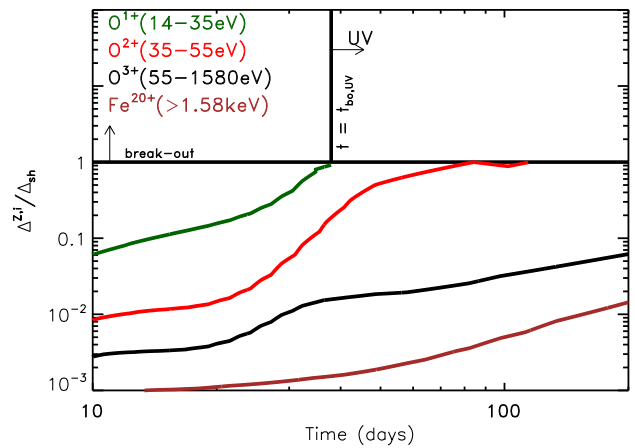}}
\caption[]{{\it \normalsize Top Panel:} Same as top panel of Figure \ref{fig:fig1}, except calculated for a pulsar with $B_{\rm d} = 10^{14}$ G and $P = 1$ ms. {\it \normalsize Bottom Panel:} Same as Figure \ref{fig:ionthick1}, but for same calculation as shown in top panel.}
\label{fig:fig2}
\end{figure}

Although ionization break-out occurs at X-ray frequencies in some cases, it is not guaranteed, even among systems capable of producing SLSNe.   Figure \ref{fig:fig2} shows a calculation for a millisecond pulsar with a magnetic field ten times stronger ($B_{\rm d} = 10^{14}$ G) than the example shown in Figure \ref{fig:fig1}.   In this case the supernova luminosity still peaks at a high value $L_{\rm SN} \sim 10^{44}$ erg s$^{-1}$ and lasts for $\gtrsim $ month, similar to the light curves of many hydrogen-poor SLSNe.  Although the O$^{1+}$ front reaches the ejecta surface near optical maximum (allowing UV break-out), the O$^{3+}$ and Fe$^{20+}$ fronts remain below the surface until much later times $\gg$ years, preventing X-rays from escaping on times of interest.  This failure to achieve break-out conditions stems from the short spin-down time of the pulsar, $t_{\rm sd} \sim 3\times 10^{4}$ s $\simeq 0.01 t_{\rm d,0}$, which results in a much lower non-thermal energy density in the nebula at times when the ejecta density has decreased sufficiently for an ionization front to form.  
 
It is in fact questionable whether break-out will occur at all in these systems on timescales of relevance, since at late times the compactness of the nebula decreases to $\ell \ll 1$.  This may terminate the formation of a pair cascade inside the nebula, which in turn could appreciably reduce the ionizing soft X-ray radiation as compared to the assumption of our model (eq.~[\ref{eq:edotsd}]) since the injected photon spectrum will now become harder ($\alpha < 1$) and extend to higher energies $h\nu \gg m_e c^{2}$.  

\begin{figure}
\subfigure{
\includegraphics[width=0.5\textwidth]{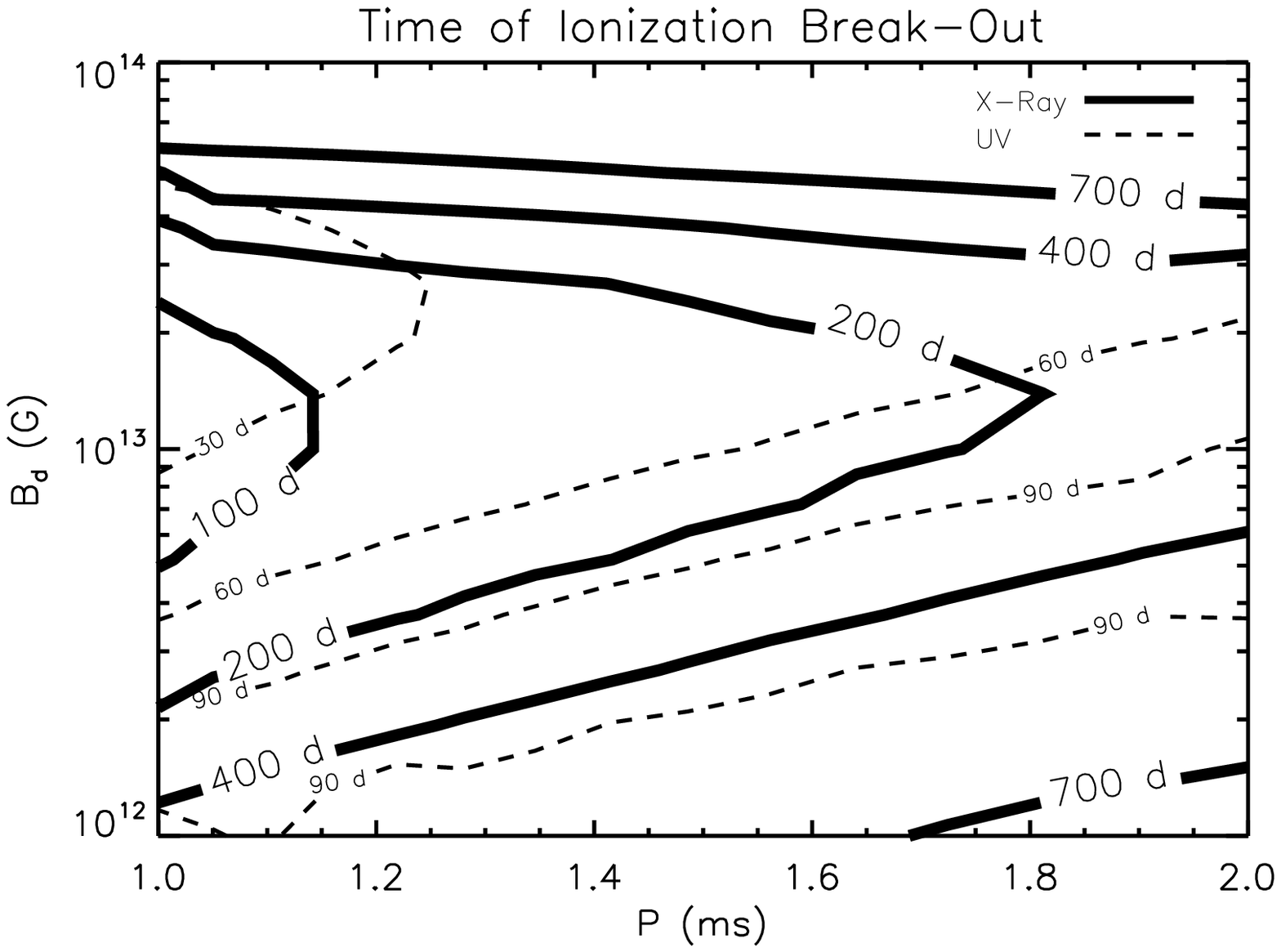}}
\subfigure{
\includegraphics[width=0.5\textwidth]{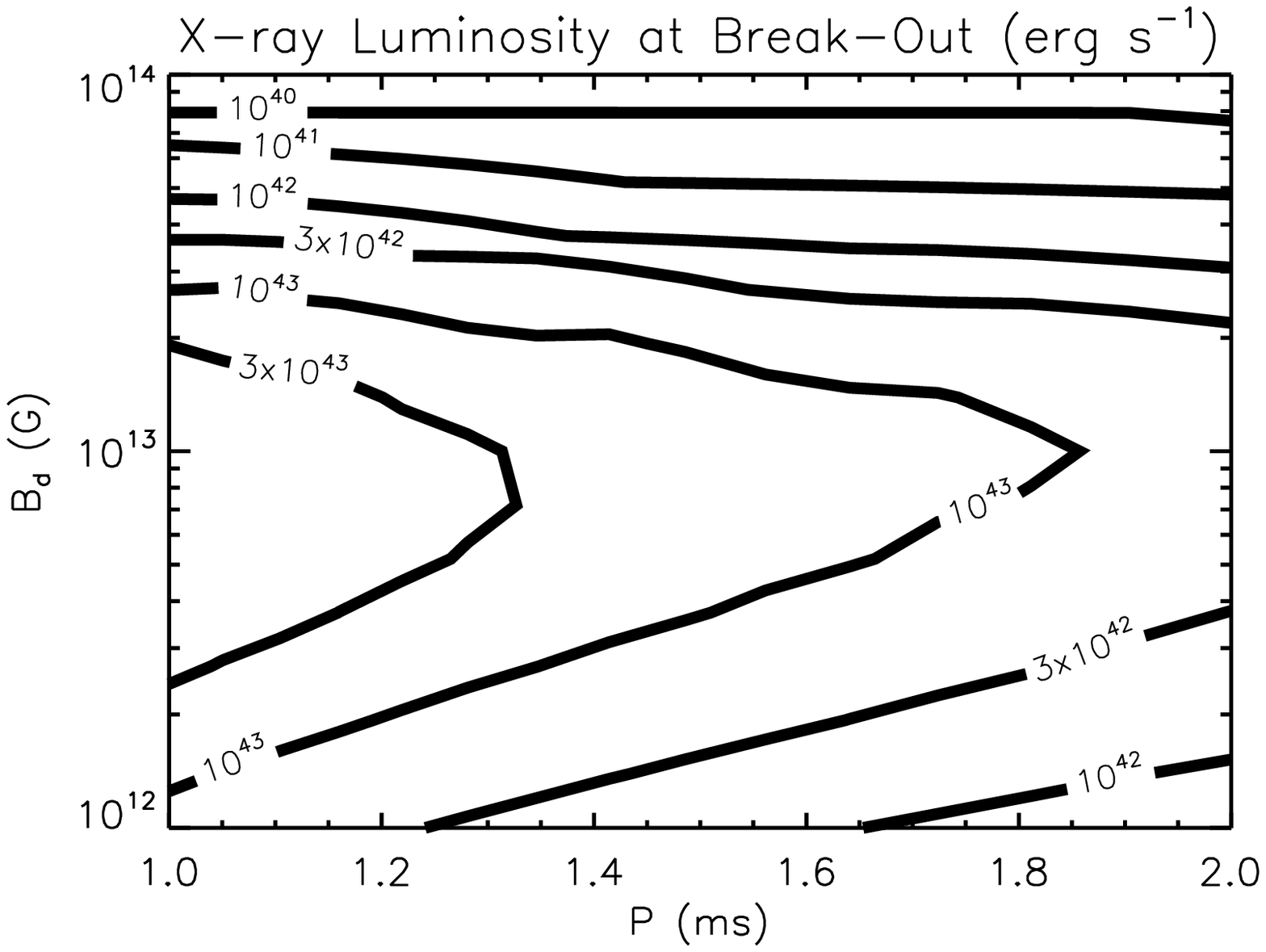}}
\caption{{\it \normalsize Top Panel:} Time of ionization break-out in the parameter space of $B_{\rm d}$ and $P$, calculated for supernova ejecta with $M_{\rm ej} = 3M_{\odot}$, $v_{\rm ej} = 10^{9}$ cm $^{-1}$, $\delta = 1$.  Thick solid lines show the break-out time at X-ray energies $\sim 0.1-1 $ keV, while dashed lines show the time of break-out at UV energies.  
{\it \normalsize Bottom Panel:} X-ray luminosity at ionization break-out, calculated for same models shown in top panel.} 
\label{fig:contour}
\end{figure}
\begin{figure}
\includegraphics[width=0.4\textwidth]{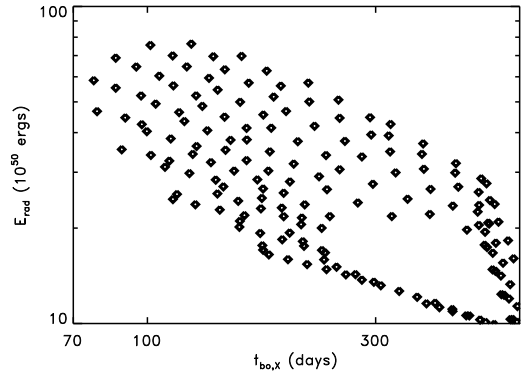}
\caption{Bolometric energy radiated by the supernova $E_{\rm rad} = \int L_{\rm SN}dt$ versus the time of X-ray ionization break-out $t_{\rm bo,X}$, from the grid of calculations used to produce Figure \ref{fig:contour}.  SLSNe with higher fluence are more likely to have an early (and luminous) X-ray break-out.} 
\label{fig:correlation}
\end{figure}
\begin{figure}
\subfigure{
\includegraphics[width=0.5\textwidth]{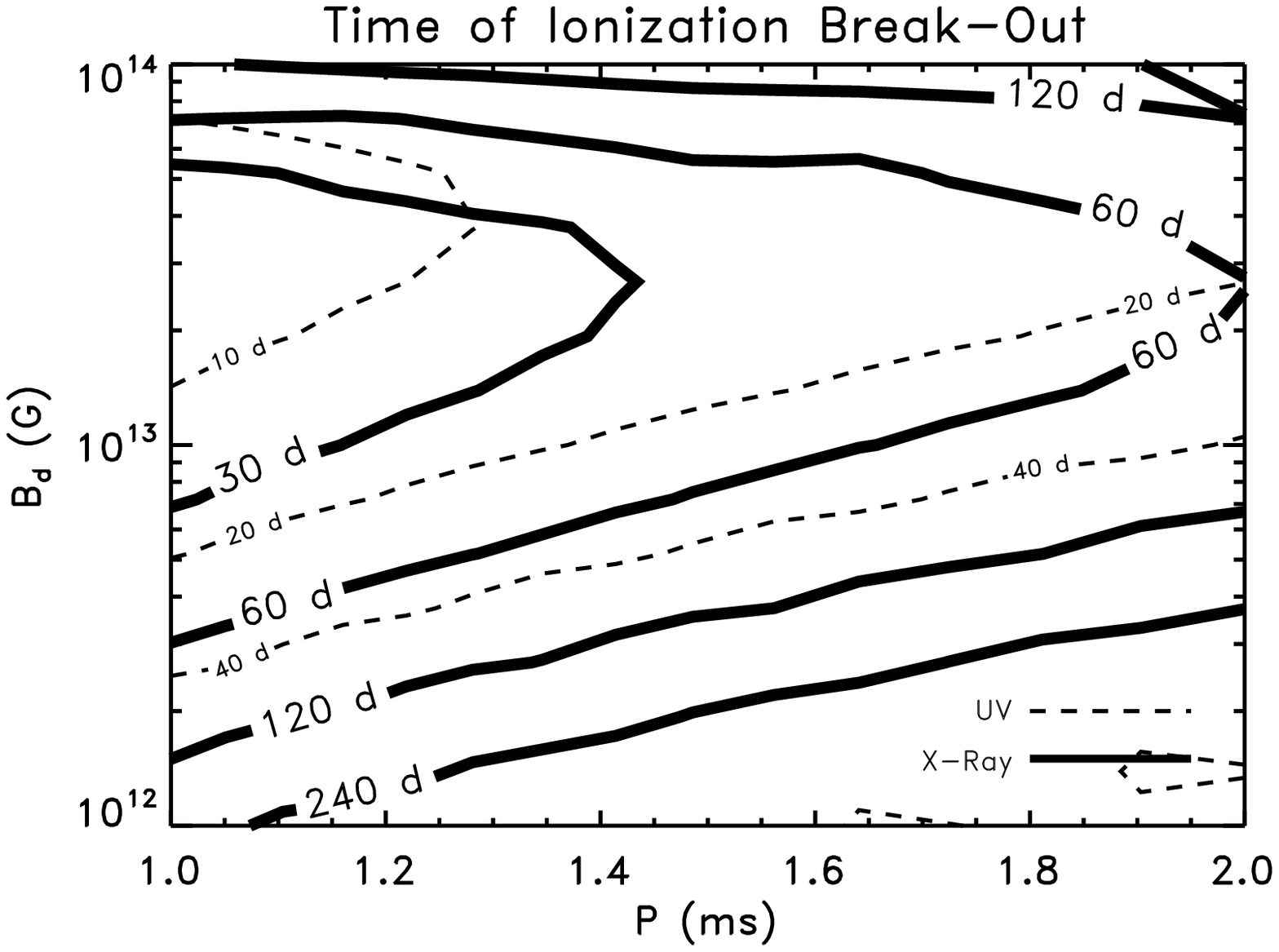}}
\subfigure{
\includegraphics[width=0.5\textwidth]{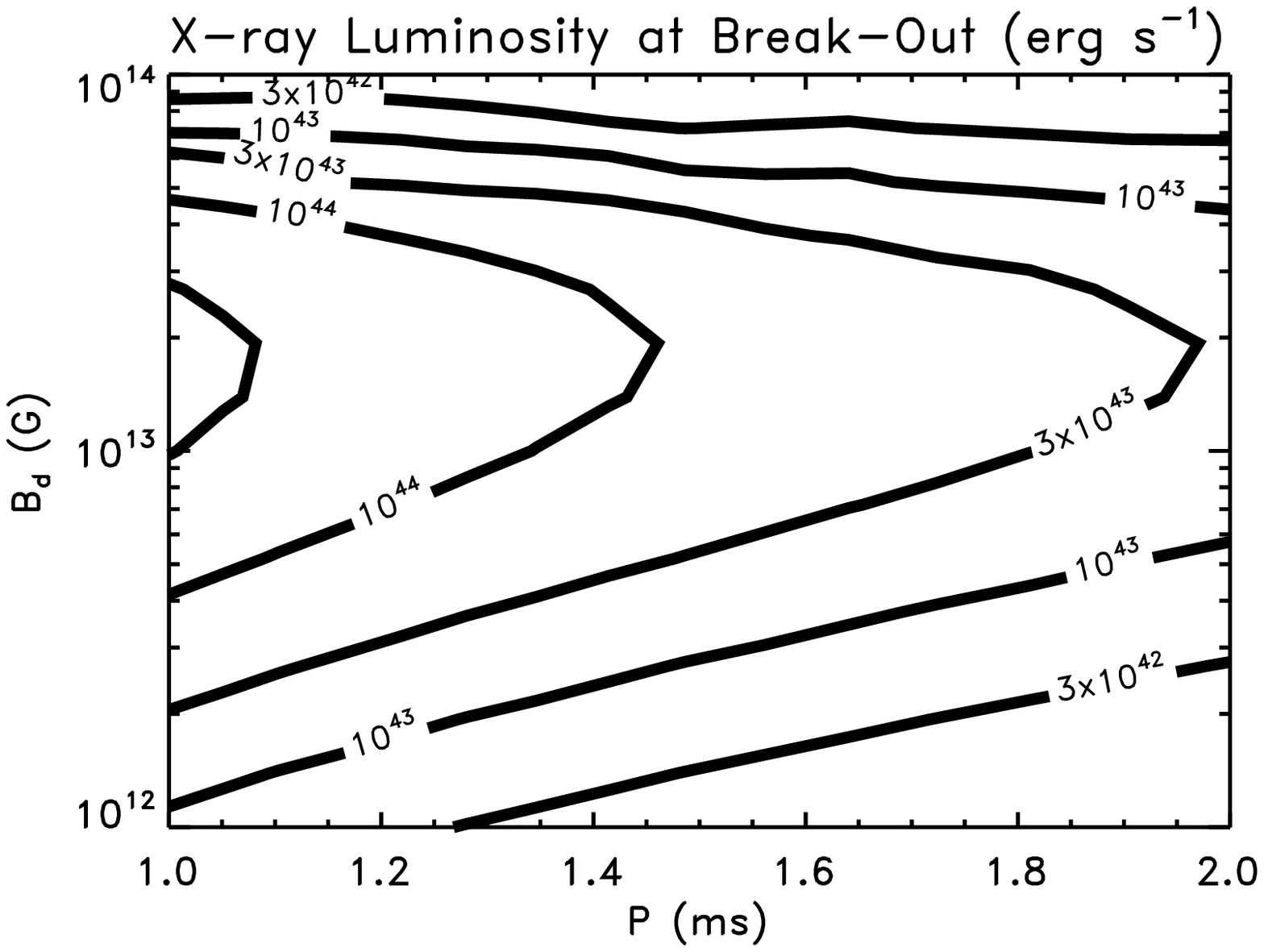}}
\caption{Same as Figure \ref{fig:contour}, but calculated for a lower ejecta mass $M_{\rm ej} = 1 M_{\odot}$.  This case may apply to  supernova with a higher total ejecta mass $M_{\rm tot} = 3M_{\odot}$, but with a 3:1 contrast between the ejecta density along the polar and equatorial direction, as would be caused by an aspherical bipolar explosion ($\S\ref{sec:aspherical}$).} 
\label{fig:contour2}
\end{figure}

Similar calculations to those shown in Figures \ref{fig:fig1}$-$\ref{fig:fig2} were performed covering a wide range of parameters, corresponding to different ejecta mass $M_3 \sim 0.3-2$; ejecta velocity $v_9 \sim 1-2$; density profile $\delta = 0-1$ and pulsar properties  ($B_{\rm d}, P$).  Our results are summarized in Figures \ref{fig:contour}$-$\ref{fig:contour2} and Table \ref{table:runs}.  Break-out is most sensitive to the ejecta mass, with a lower(higher) value of $M_{\rm ej}$ producing earlier(later) break-out of higher(lower) X-ray luminosity.  

Figure \ref{fig:contour} shows contours of break-out time as a function of $P$ and $B_{\rm d}$ for fiducial parameters $M_3 = v_9 = \delta = 1.$  The O$^{1+}$ layer reaches the ejecta surface promptly across a wide range of parameter space, making UV break-out relatively ubiquitous among SLSNe.  However, the O$^{3+}$ layer only promptly reaches the surface, allowing X-ray break-out, across a fairly narrow range of pulsar properties, requiring very rapid rotation ($P \lesssim 2$ ms) and intermediate values of $B_{\rm d} \sim 3\times 10^{12}-3\times 10^{13}$ G for the fiducial ejecta mass $M_{\rm ej} = 3 M_{\odot}$ (Fig.~\ref{fig:contour},\ref{fig:contour2}).  Not coincidentally, this region corresponds to that capable of producing the highest supernova fluence, as shown by the correlation in Figure \ref{fig:correlation} between the bolometric radiated energy of the supernova and the time of X-ray break-out among the models used to construct Figure \ref{fig:contour}.  Both ionization break-out and high optical fluence favor pulsars with the highest possible rotation rate ($P \simeq 1$ ms) since this provides the largest energy reservoir.  A magnetic field $B_{\rm d} \sim 10^{13}$ G is also optimal since it renders the spin-down time $t_{\rm sd}$ (for $P = 1$ ms) comparable to the photon diffusion time at optical peak $t_{\rm d,0}$.

\begin{table*}
\begin{scriptsize}
\begin{center}
\vspace{0.05 in}\caption{Properties of Ionization Break-Out from Millisecond PWNe}
\label{table:runs}

\begin{tabular}{ccccccccccc}
\hline \hline
\\
\multicolumn{1}{c}{B$_{\rm d}$} &
\multicolumn{1}{c}{P} &
\multicolumn{1}{c}{M$_{\rm ej}$} &
\multicolumn{1}{c}{v$_{\rm ej}$} &
\multicolumn{1}{c}{$\delta$} & 
\multicolumn{1}{c}{$t_{\rm cr}^{(a)}$} & 
\multicolumn{1}{c}{$E_{\rm rad}^{(b)}$} & 
\multicolumn{1}{c}{t$_{\rm bo}^{\rm UV}$$^{(c)}$} &
\multicolumn{1}{c}{t$_{\rm bo}^{\rm X}$$^{(d)}$} &
\multicolumn{1}{c}{$\ell$($t_{\rm bo}^{\rm X})^{(e)}$} &
\multicolumn{1}{c}{L$_{\rm X,bo}^{(f)}$} \\
\hline
G & ms & $M_{\odot}$ & 10$^{9}$ cm s$^{-1}$ & - & {\rm days} & $10^{50}$ ergs & {\rm days} & days & - & $10^{43}$ erg s$^{-1}$ \\
\hline
\\
$10^{13}$ & 1 & 3 & 1 & 1 & 7 & 58 & 27 & 77 & 1.6 &  8 \\
- & - & - & 2 & - & 21 & 59 & 27 & 59 & 1.9 &  12 \\
- & - & 1 & - & - & 4 & 54 & 12 & 23 & 19 &  31 \\
- & - & 6 & - & - & 12 & 56 & 49 & 213 & 0.12 &  1.7 \\
- & - & 3 & - & 0 & 7 & 65 & 33 & 83 & 1.0 &  7 \\
$3\times 10^{13}$ & - & - & - & 1 & 3 & 23 & 21 & 129 & 0.06 &  0.65 \\
$3\times 10^{12}$ & - & - & - & - & 41 & 73 & 70 & 141 & 0.43 & 4.1 \\
$10^{13}$ & 2 & - & - & - & 68 & 22 & 94 & 240 & 0.05 & 0.70 \\
 \\
  \\
\hline
\hline
\end{tabular}
\end{center}
\end{scriptsize}
$^{(a)}$Time at which shock driven by nebular pressure reaches front of ejecta.  $^{(b)}$ Total radiated optical fluence.  $^{(c)}$ Time after which the O$^{1+}$ layer reaches the ejecta surface, allowing UV photons to escape the ejecta (UV break-out).  $^{(d)}$ Time after which the O$^{3+}$ layer reaches the ejecta surface, allowing $\sim$ 0.1-1 keV X-ray photons to escape (X-ray break-out).  $^{(e)}$ Compactness of nebula at time of X-ray break-out. $^{(f)}$ X-ray luminosity immediately after break-out (eq.~[\ref{eq:LXbo}]).
\end{table*}

\section{Discussion}
\label{sec:discussion}

\subsection{Implications for X-ray Emission from SLSNe}
\label{sec:SLSNeX}

The electron scattering diffusion timescale at the time of X-ray break-out, $t_{\rm d,bo}$, is generally shorter than the break-out time itself, $t_{\rm bo}$.  The observed X-ray luminosity immediately after break-out $L_{\rm X,bo}$ is thus set by the pulsar luminosity in the $\sim 0.1-1$ keV frequency range of relevance:
\begin{eqnarray}
L_{\rm X,bo} \sim \left.\frac{L_{\rm sd}}{7}\right|_{t = t_{\rm bo}} \sim 10^{45}{\rm erg\,s^{-1}}B_{13}^{-2}M_{3}^{-1}v_9\left(\frac{t_{\rm bo}}{t_{\rm d,0}}\right)^{-2}
\label{eq:LXbo}
\end{eqnarray}
where the factor of 7 accounts for the non-thermal luminosity of the cooling pairs being divided equally per decade in frequency between the thermal bath $h\nu \sim 1$ eV and the pair production cut-off $h\nu \sim 1 $ MeV (cf.~eq.~[\ref{eq:edotsd}]), and in the second equality we have assumed that $t_{\rm bo} \gtrsim t_{\rm sd}$, as is usually satisfied. The bottom panel of Figure \ref{fig:contour} shows contours of $L_{\rm X,bo}$ calculated directly from our grid of numerical models, showing reasonable agreement with the estimate in equation (\ref{eq:LXbo}).  

The rise time of the X-ray emission is set by the time required to achieve ionization break-out across the frequency band of interest, which may be comparable to the break-out time itself, due to the order unity difference in the penetration depth (and hence break-out time) of photons with different frequency across the band (Fig.~\ref{fig:iondepth}).  in principle the rise time at a given frequency could be much more rapid, but in practice there will be a finite delay since it is unlikely that the conditions for break-out will be achieved simultaneously across the entire ejecta surface.  The peak luminosity $L_{\rm X,bo}$ is maintained for a characteristic timescale 
\begin{eqnarray}
\Delta t_{\rm X} \sim t_{\rm bo},
\label{eq:delta_X}
\end{eqnarray}
also of order the break-out time, after which the X-ray light curve decays with the pulsar luminosity as $L_{\rm X} \simeq L_{\rm X,bo}(t/t_{\rm bo})^{-2}$.

\begin{figure*}
\includegraphics[width=0.5\textwidth]{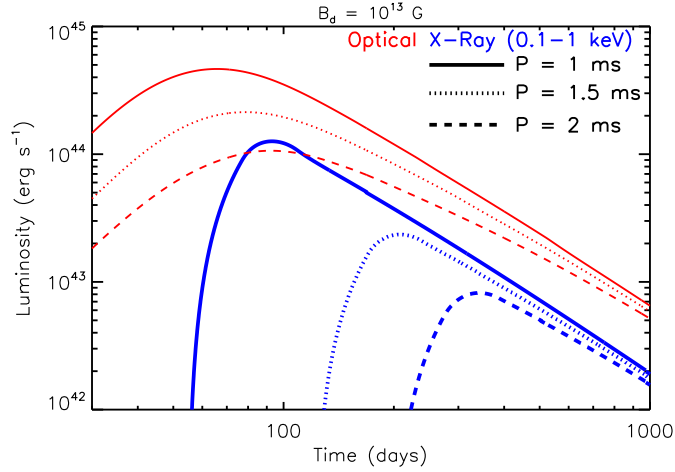}
\caption{Optical and X-ray light curves for pulsar-powered SLSNe, calculated for a pulsar with dipole field $B_{\rm d} =  10^{13}$ G and for three different initial spin periods $P$ = 1, 1.5, and 2 ms, as marked.  The ejecta has mass $M_{\rm ej} = 3M_{\odot}$, initial velocity $v_{\rm ej} = 10^{9}$ cm s$^{-1}$, and density profile $\delta = 1$.   The rise of the X-ray light curve and its shape near the peak is somewhat schematic (see text).} 
\label{fig:LC}
\end{figure*}

Thus, in a typical case for which $t_{\rm bo} \sim$ few months $\sim$ few $t_{\rm d,0}$ (e.g.~Fig.~\ref{fig:ionthick1}) we have $L_{\rm X} \sim 10^{43}-10^{44}$ erg s$^{-1}$ and $\Delta t_{\rm X} \sim$ months, resulting in a total radiated X-ray energy up to $\sim 10^{50}-10^{51}$ ergs.  
Figure \ref{fig:LC} shows a schematic diagram of the optical and X-ray light curves for a magnetar with $B_{\rm d} = 10^{13}$ G and several initial spin periods.   The rise of the X-ray light curve and its shape near the peak are plotted somewhat schematically, because in reality these will depend on the detailed evolution of the break-out process across the observing band, and how that the radiation field in the nebula relaxes from its state prior to break-out to the new balance achieved once radiative losses become important.  The late-time X-ray light curve may also deviate from the predicted $L_{\rm X} \propto t^{-2}$ decay if the photon spectrum injected by cooling pairs from the pulsar changes from the flat distribution $J_{\nu} \propto \nu^{-1}$ assumed in this work once the pair cascade ceases.

Luminous X-rays were detected from the hydrogen-poor superluminous supernovae SCP 06F6 approximately $t \sim 75$ days (rest frame) after the initial discovery (\citealt{Levan+13}).  The timescale of this emission, approximately one month after the optical peak of the supernova, is consistent with the X-rays being the result of ionization break-out from a pulsar wind nebula as described here (Fig.~\ref{fig:ionthick1}).  Ionization break-out is indeed favored to occur after the optical peak because the growing width of the ionization front is a competition between the ionizing X-ray flux (which is greatest near the optical peak when the nebula energy is highest) and the rate of recombination, which decreases as the ejecta expands.   

The 0.2$-$2 keV luminosity of SCP 06F6, $L_{\rm X} \sim 6\times 10^{44} - 2\times 10^{45}$ erg s$^{-1}$, is consistent with the luminosity expected from ionization break-out for $t_{\rm bo} \sim 2t_{\rm d,0} \sim$ 75 days (eq.~[\ref{eq:LXbo}]), especially given the range of theoretical and observational\footnote{The derived range of luminosities depends on whether the X-ray spectrum is fit to a thermal or an (unattenuated) power-law model, respectively.  The photon statistics were insufficient to discriminate between these models.} uncertainties and taking into account the possible enhancement to the X-ray luminosity if break-out occurred along the poles of the ejecta, assuming the latter pointed towards our line of sight (see $\S\ref{sec:aspherical}$ below).  

The rise time $\sim 60$ days and peak luminosity $L_{\rm SN} \sim 5\times 10^{44}$ erg s$^{-1}$ of the optical light curve of our fiducial model producing an X-ray break-out (Fig.~\ref{fig:fig1}) match those of SCP 06F6 to within a factor $\lesssim 2$ (\citealt{Chatzopoulos+13}).  It would not be coincidental that SCP 06F6 has one of the highest radiated optical energy of any SLSN Type I since optical fluence is correlated with likelihood of X-ray break-out (Fig.~\ref{fig:correlation}).  A puzzling feature of the X-ray emission from SCP 06F6 is the subsequent non-detection by {\it Chandra} just $ t\sim 40$ days after the initial X-ray detection, placing an upper limit on the flux a factor $\sim 2.5-8$ times lower than its initial value.  This is marginally consistent with the predicted decay $L_{\rm X} \propto (t/t_{\rm bo})^{-2} \sim 2.5$ (eq.~[\ref{eq:delta_X}]) in the X-ray flux after ionization break-out.  

Other than SCP 06F6, X-ray emission has not been detected from other SLSN-I.  When available, upper limits on the luminosity are in the range $L_{\rm X} \lesssim 10^{43}-10^{44}$ erg s$^{-1}$ (see Fig.~3 of Levan et al.~2013).  Since many of these are below the range expected for ionization break-out (eq.~[\ref{eq:LXbo}]), we may conclude that X-ray break-out did not occur in these events, at least at the time of observations.  This in itself does not disfavor an engine-powered origin for these events, because the conditions required to get X-ray ionization break-out are more restrictive than those required to get a SLSNe (Fig.~\ref{fig:contour}).  Furthermore, most of the upper limits are on timescales which may be too early ($\lesssim$ 70 days) or too late ($\gtrsim 300$ days) to detect the break-out phase, even if it occurred. 

\subsection{Implications for Blue Colors of SLSN-I}
\label{sec:UV}

Across a relatively wide parameter space, the ionization front of O$^{1+}$ reaches the ejecta surface with relative ease (Fig.~\ref{fig:contour}; Table \ref{table:runs}).  Thus, unlike X-rays, the escape of UV radiation from the nebula starting near the time of optical peak (Table \ref{table:runs}), may be a robust feature of the pulsar-powered model for SLSNe.  

Interestingly, SLSN-I are observed to have an especially blue spectral energy distribution (Quimby et al. 2011).  This could be due to a low iron abundance as compared to normal Type I supernovae (Quimby et al.~2011), or it could be due to nebular ionization of the ejecta as discussed here.  Future work is required to better quantify and test this idea.  An important next step is to include self-consistently the effects of line opacity, which itself will be modified by the ionization of valence states of iron or other elements in the ejecta.

\subsection{Effects of Aspherical or Clumpy Supernova Ejecta}

\label{sec:aspherical}

Figure \ref{fig:contour2} shows the results of a grid of calculations otherwise identical to those in Figure \ref{fig:contour}, but assuming a smaller ejecta mass $M_{\rm ej} = 1M_{\odot}$.  In this case the prompt break-out extends across a much wider parameter space ($P$, $B_{\rm d}$), on a typically much shorter timescale $\lesssim 30-100$ days.  Break-out is easier to achieve when $M_{\rm ej}$ is lower due to the lower electron density (slower recombination) and higher ejecta velocity achieved for a given pulsar rotational energy (eq.~[\ref{eq:vejf}]).  

Although such a low ejecta mass $\sim M_{\odot}$ is probably unphysical, at least in the case of a supernova that leaves behind a neutron star remnant, this calculation may nevertheless be physically relevant to the case of aspherical (e.g.~bipolar) ejecta.  A 3:1 density contrast between the equator and poles of the ejecta would, for instance, reduce the {\it effective} ejecta mass for a polar break-out to $\sim 1M_{\odot}$, even for a total mass of $3M_{\odot}$.  Strongly aspherical break-out could also  impart a moderate directional dependence to the X-ray flux, such that lines of sight along the pole (equator) receive a larger (smaller) fraction of the break-out luminosity.  Significant bipolar asymmetry of the ejecta could be expected given the extremely rapid rotation of the progenitor star, and the rotationally-powered explosion, required to birth a millisecond pulsar in the first place.  

Given the characteristic densities and temperatures of the ionized ejecta, it is also possible that this medium will be susceptible to thermal instabilities, as occurs in e.g.~the broad line regions of quasars (e.g.~\citealt{Krolik+81}; \citealt{Ferland&Elitzur84}).  Thermal instability could lead to clumping, the enhanced porosity of which might give rise conditions more favorable for the escape of ionizing photons than predicted by our homogeneous model.  

\section{Conclusions}
\label{sec:conclusions}

Energy injection from a newly-born millisecond pulsar is a promising explanation for powering the most luminous supernovae yet discovered.  However, degeneracies of the model make it difficult to test with confidence based on the optical light curve alone and to rule out other possible energy sources, such as optically thick circumstellar interaction.  The model also makes several implicit and possibly unjustified assumptions, such as requiring that the bulk energy of the wind be deposited as thermal energy behind the supernova shock, producing a particle-dominated pulsar wind nebula.  This required dissipation may be justified physically by global instabilities and magnetic reconnection in the nebula (e.g.~\citealt{Porth&Komissarov13}), but this is by no means assured.  Even assuming that such dissipation occurs, no calculation had yet addressed how the relativistic pairs to which the wind initially endows this energy are able to transfer it to the thermal bath of the nebula, which ultimately powers much of the observed optical luminosity.  

In this work we {\it assume} that such dissipation occurs, and explore its observational consequences.  If pairs are injected with sufficiently high Lorentz factors, their cooling necessarily produces a spectrum of hard X-rays inside the nebula due to Compton up-scattering and synchrotron radiation.  The ejecta is highly opaque to X-rays at early times, and efficient absorption by the ejecta walls deposits a sizable fraction of the pulsar power into the nebula thermal bath (Fig.~\ref{fig:fig1}).  However, due to the ongoing dissipation and partial reflection of X-rays by the ejecta walls, an energetic hard radiation field remains pent up inside the expanding nebula (Fig.~\ref{fig:spectra}).  If the spin-down timescale of the pulsar is comparable to the diffusion time of the ejecta, then this radiation can be sufficiently luminous to re-ionize the ejecta in its entirety.  This typically occurs on timescales of $\sim$ months after the beginning of the explosion, allowing $\gtrsim$ 0.1 keV photons to escape (Fig.~\ref{fig:ionthick1}).  The X-ray luminosity at break-out is necessarily very high $\gtrsim 10^{43}$ erg s$^{-1}$ (eq.~[\ref{eq:LXbo}]; since otherwise it would not be sufficient to achieve break-out in the first place), rendering it detectable even out to the large distances $z \sim 0.1-1$ that characterize most SLSNe.

The break-out X-ray spectrum is a power-law with a possible cut-off at $h\nu \sim$ few keV (rest frame), due to absorption by higher $Z$ elements (e.g.~iron), which may not be fully ionized throughout the ejecta.  This cut-off is predicted to move to higher energies with time, as additional elements become fully ionized.  Additional work is required to better quantify the time-dependent shape and extent of the cut-off, but this unique soft-to-hard spectral evolution may allow future observations to distinguish between this X-ray emission and that produced by circumstellar interaction (e.g.~\citealt{Chevalier&Irwin11}; \citealt{Ofek+13}; \citealt{Pan+13}).   Another potentially important distinguishing feature is the continuum X-ray spectrum $\nu F_{\nu} \propto \nu^{\alpha}$, which is approximately flat after ionization break-out ($\alpha \sim 0-0.5$) in pulsar-driven models, as compared to the steeper spectrum $\alpha \sim 1$ of thermal bremsstrahlung emission due to X-ray luminous circumstellar interaction (e.g.~\citealt{Pan+13}).  Given that X-ray break-out is often marginal, and may not be achieved simultaneously at all frequencies, the X-ray spectrum just after ionization break-out may also be characterized by absorption lines of highly ionized states of oxygen or other abundant elements in the ejecta. 


The conditions for prompt X-ray break-out are not easily achieved (Figs.~\ref{fig:fig2}, \ref{fig:contour}), making it unlikely that all SLSNe-I are accompanied by bright X-ray emission (Fig.~\ref{fig:contour}).  However, we emphasize that X-ray observations of a SLSN-I {\it should not cease if no detection is made near optical maximum, since the onset of X-ray ionization break-out is often delayed with respect to the supernova peak.}  Observations should also be densely sampled in time, since the duration of peak X-ray brightness can be as brief as a few weeks.  A rough trend is that SLSNe with higher optical fluence are also more likely to produce a successful and luminous X-ray break-out (Fig.~\ref{fig:correlation}), which may help prioritize follow-up.

Much of the physics discussed here could in principle apply also to SLSNe powered by accretion onto a newly-formed black hole (\citealt{Quataert&Kasen12}; \citealt{Dexter&Kasen12}), provided that the power input comes from a black hole jet that was itself comprised chiefly of electron/positron pairs.  Models in which the supernova is powered by outflows from the black hole accretion disk would not directly apply if, however, the outflow is too baryon-loaded to produce the requisite X-ray nebula.

Recent work by \citet{Kotera+13} also explored the implications for early X-ray emission from millisecond pulsars, from which they predict X-ray luminosities up to $L_{\rm X} \sim 10^{41}-10^{42}$ erg s$^{-1}$ on timescales of months to years.  The main differences between the work of Kotera et al.~and our model is the source of the X-ray emission.   Kotera et al.~do not include the effect of the pair cascade or ejecta ionization discussed here, so the predicted emission in their models presumably originates from hard X-rays ($\gtrsim 10$ keV) which escape from the nebula above the bound-free cut-off of iron.  We instead focused on softer X-ray emission here because ionization break-out allows for much earlier emission at much higher luminosities, and because the emission from SCP 06F6 was in the energy range $\sim 0.1-2$ keV.  

Many aspects of the model developed here are simplified and will be improved in future work.  First, a self-consistent model of the high energy processes occurring in the nebula (pair creation, synchrotron radiation, and Compton scattering) is necessary to account for the changing nature of the radiation field produced by the cooling pairs, especially as the nebula compactness drops below unity at late times and the saturated pair cascade ceases.  Second, our model for the thermal evolution of the nebula neglects energy loss at high frequencies, despite the obvious importance of such losses after ionization break-out.  The process by which the nebula exchanges energy with the surrounding ejecta should also be better quantified beyond the single diffusion timescale treatment presented here.   Such details are necessary to produce a detailed comparison between our model and the observed optical light curves of SLSNe.  Finally, we have taken a highly simplified approach in calculating the time-dependent ionization structure of the ejecta.  Our `one zone' model for the ionized layer(s) should be replaced by a self-consistent solution for the radiation field, temperature, and ionization fractions as a function of depth through the ejecta, including a wider range of elements and a more detailed treatment of the bound-bound (line) opacity (e.g.~\citealt{Ross&Fabian05}).  This is necessary to produce more accurate estimates of the UV/X-ray spectrum at the time of break-out and to assess quantitatively whether nebular ionization contributes to the blue/UV SED of SLSN-I ($\S\ref{sec:UV}$).

\section*{Acknowledgments}
BDM acknowledges helpful conversations with Roger Chevalier and Andrew Levan.  BDM acknowledges support from the Columbia University Department of Physics.  IV, RH, and AMB acknowledge support from NSF grant AST-1008334.

\appendix

\section{Condition for Ionization Break-Out of Hydrogen-Like Species}
\label{sec:hlike}

\begin{figure}
\subfigure{
\includegraphics[width=0.5\textwidth]{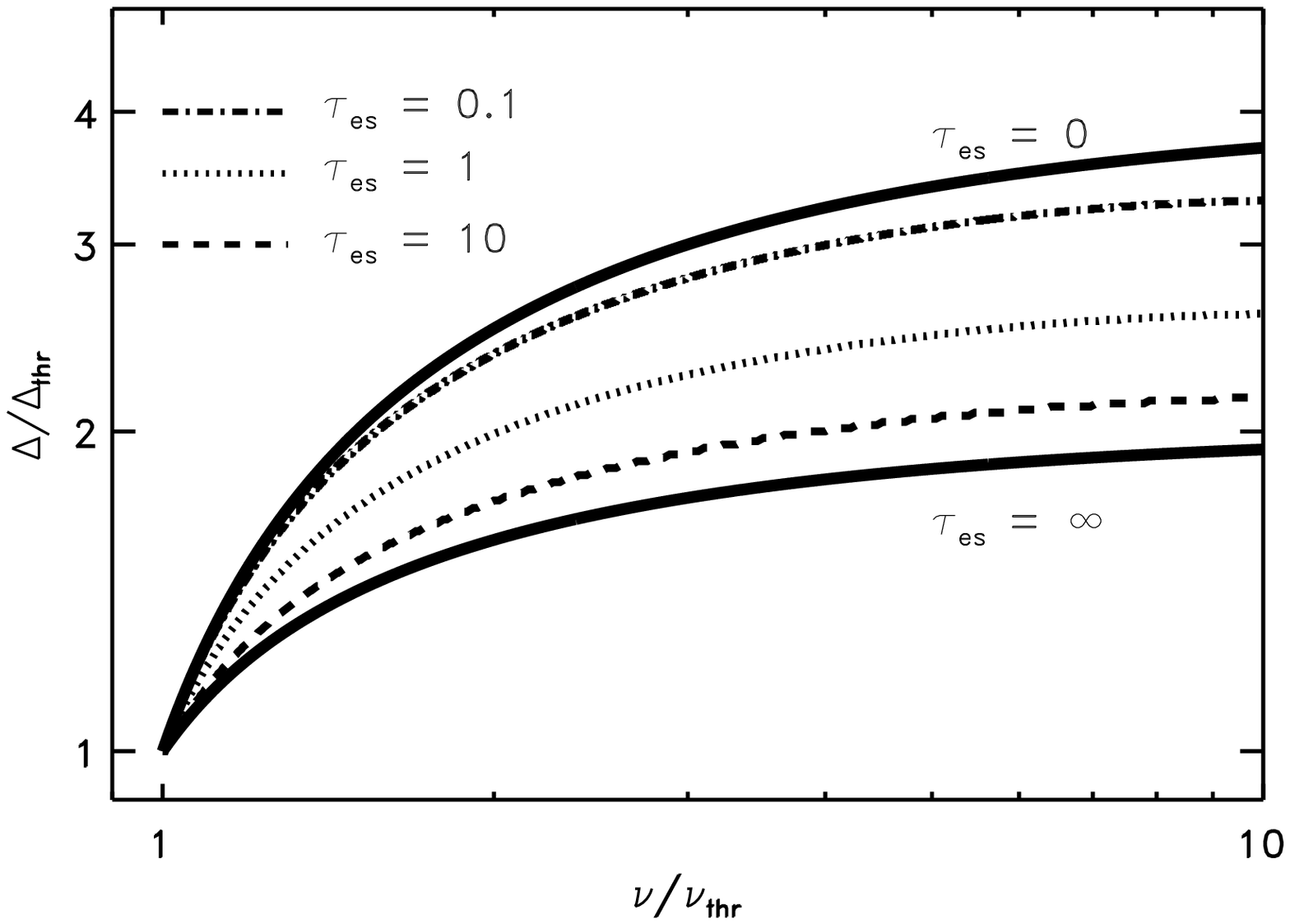}}
\subfigure{
\includegraphics[width=0.5\textwidth]{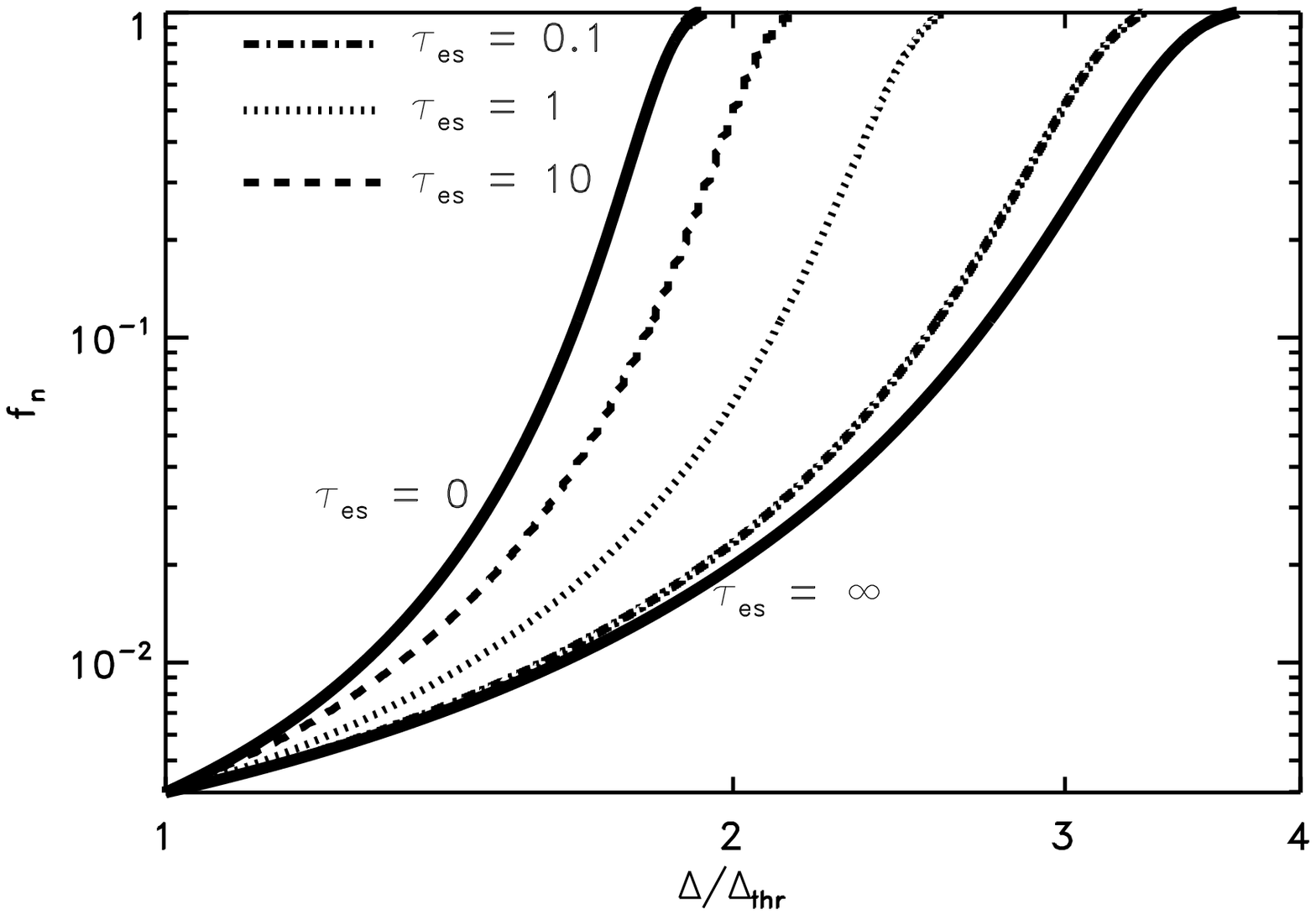}}
\caption{{\normalsize \it Top Panel:} Ionization depth $\Delta$ of photons of frequency $\nu > \nu_{\rm thr}$ as compared to the depth $\Delta_{\rm thr}$ of photons at the ionization threshold frequency $\nu = \nu_{\rm thr}$, calculated from equation (\ref{eq:tau2}) for different values of the electron scattering optical depth $\tau_{\rm es}$ at $\Delta_{\rm thr}$. {\it \normalsize Bottom Panel:} Neutral fraction $f_{\rm n}$ as a function of depth through the ionized layer for the same cases shown in the top panel, calculated assuming a neutral fraction $f_{\rm n} = 4\times 10^{-3}$ at depth $\Delta_{\rm thr}$.} 
\label{fig:iondepth}
\end{figure}

Here we describe the ionization structure of the ejecta in greater detail and approximate analytically the requisite conditions for ionization break-out.  For simplicity we focus on the ionization of hydrogen-like atomic species and assume that each element $Z$ sets the penetration depth of photons with frequency above the ionization threshold energy
\be
h \nu_{\rm thr} = 13.6Z^{2} {\rm eV}.
\ee  
In the actual case, all ionization states must be considered, but only those of the most abundant elements (e.g.~helium, oxygen, iron) are relevant in setting the ionization depth.  

The characteristic penetration depth $\Delta$ of a photon of frequency $\nu > \nu_{\rm thr}$ is achieved at an effective optical depth of unity (eq.~[\ref{eq:tau1}]), as defined by the integral condition
\be
\int_{0}^{s = \Delta}\rho_{\rm sh}\kappa_{\rm abs,\nu}(s')\left[1 + \kappa_{\rm es}\rho_{\rm sh}s'\right]ds' = 1,
\label{eq:tau2}
\ee
where $s$ is the depth through the shocked layer and $\kappa_{\rm abs,\nu}(s) = f_n(s) X_{A}\sigma_{\rm bf,\nu}/A m_p$ is the bound-free opacity (eq.~[\ref{eq:kappa}]), where the cross section for hydrogen-like species is approximately given by (e.g.~\citealt{Osterbrock&Ferland06})
\be
\sigma_{\rm bf}(\nu) = \frac{8\times 10^{-18}{\rm cm^{2}}}{Z^{2}}\left(\frac{\nu_{\rm thr}}{\nu}\right)^{3},
\label{eq:sigmabf}
\ee
The neutral fraction $f_{\rm n} \ll 1$ (eq.~[\ref{eq:fn}]) 
\begin{eqnarray}
f_n(s) &\approx& \frac{\alpha_{\rm rec} n_e V_{\rm n}}{c}\left(\int_{\nu_{\rm min}(s)}^{\nu_{\rm max}}\frac{E_{\rm nth,\nu}}{h\nu}\sigma_{\rm bf}(\nu)d\nu\right)^{-1}
\label{eq:fna}
\end{eqnarray}
varies with depth $s$ because only photons that have not yet been absorbed by depth $s$ (those with $\nu > \nu_{\rm min}(s)$) contribute to the local photoionization rate, where $\nu_{\rm min}$ is the frequency determined by condition (\ref{eq:tau2}) for $\Delta = s$.  Here $\nu_{\rm max}$ is the maximum frequency of the band, which is usually the threshold frequency of the next relevant ionization state and $\alpha_{\rm rec}$ is the the [type B] recombination coefficient, which is approximately given by (e.g.~\citealt{Osterbrock&Ferland06})
\be
\alpha_{\rm rec} = 4\times 10^{-14}Z^{2}T_{5}^{-0.8}{\rm cm^{3}s^{-1}},
\label{eq:alpharec}
\ee
where $T_{\rm C} \simeq 10^{5}T_{5}K$ is the ejecta temperature, normalized to a value, $10^{5}$ K, characteristic of the Compton temperature of electrons in the ionized layer near the time of ionization break-out (see discussion surrounding eq.~[\ref{eq:numax}]).  

For photons at the ionization threshold frequency $\nu_{\rm thr}$, the penetration depth can be derived from equation (\ref{eq:tau2}) assuming a constant opacity $\kappa_{\rm abs,\nu}$ with depth calculated using the unattenuated nebular spectrum (i.e. $\nu_{\rm min} = \nu_{\rm thr}$ in equation \ref{eq:fna}):
\be
\frac{\Delta_{\rm thr}}{\Delta_{\rm sh}} \simeq \frac{\sqrt{1 + 4 \eta_{\rm thr}^{-1}} - 1}{2 \tau_{\rm es}^{\rm sh}},
\label{eq:dth}
\ee
where $\Delta_{\rm sh}$ and $\tau_{\rm es}^{\rm sh} = \rho_{\rm ej}\kappa_{\rm es}\Delta_{\rm sh}$ are the total thickness and scattering optical depth of the shocked ejecta, respectively, and
\begin{eqnarray}
\eta_{\rm thr} &\equiv& \left.\frac{\kappa_{\rm abs,\nu}}{\kappa_{\rm es}}\right|_{\nu = \nu_{\rm thr}} \simeq \frac{4X_{A}\alpha_{\rm rec}n_{e}V_{\rm ej}h \nu_{\rm thr}}{A \kappa_{\rm es} m_p c E_{\rm nth,\nu_{\rm thr}}\nu_{\rm thr}} \approx \frac{28 X_{A}\alpha_{\rm rec}M_{\rm ej} h \nu_{\rm thr} (1 - \mathcal{A}_{\nu})}{A \kappa_{\rm es} m_p^{2} v_{\rm ej} L_{\rm sd}t } \nonumber \\
&\approx& 0.67(1-\mathcal{A}_{\nu})\left(\frac{L_{\rm sd} t}{10^{52}{\rm erg}}\right)^{-1}M_{3}v_9^{-1}\left(\frac{X_{A}}{0.1}\right)T_{5}^{-0.8}Z_8^{3}
\label{eq:etaA}
\end{eqnarray}
is the ratio of absorptive and scattering opacity at $\nu = \nu_{\rm thr}$ (eq.~[\ref{eq:eta}]), where $Z_{8} \equiv Z/8$.  In equation (\ref{eq:etaA}) we estimate the electron density as $n_e \approx M_{\rm ej}/2m_p V_{\rm ej}$ and calculate the ionization rate assuming $E_{\rm nth,\nu} \propto \nu^{-1}$, $\nu_{\rm min} = \nu_{\rm thr}$, and $\nu_{\rm max} \rightarrow \infty$.  We also approximate the photon energy of the nebula $E_{\rm nth,\nu_{\rm thr}}$ at frequency $\nu_{\rm thr}$ by assuming a balance between energy injection from the pulsar and losses due to absorption by the ejecta walls in equation (\ref{eq:photon_evo}):
\be
\dot{E}_{\rm sd,\nu} \approx (1 - \mathcal{A}_{\nu})\frac{E_{\rm nth,\nu}}{t_{\rm d}^{\rm n}} \,\,\,\,\Rightarrow\,\,\,\, E_{\rm nth, \nu_{\rm thr}}\nu_{\rm thr} \approx \frac{L_{\rm sd}t}{14(1-\mathcal{A}_{\nu})}\frac{v_{\rm ej}}{c}
\label{eq:edotsdeq}
\ee
where $\mathcal{A}_{\nu}$ is the albedo of the ionized layer at $\nu_{\rm thr}$; $\dot{E}_{\rm sd,\nu_{\rm thr}}\nu_{\rm thr} \approx L_{\rm sd}/14$ is the rate of energy injection by the pulsar (eq.~[\ref{eq:edotsd}]), and $t_{\rm d}^{\rm n} \sim R_{\rm n}/c \approx R_{\rm ej}/c$ (eq.~[\ref{eq:tdiffn}]) is the photon crossing time of the nebula.  

Equation (\ref{eq:dth}) provides the penetration depth of photons at the threshold frequency $\nu_{\rm thr}$, but higher energy photons penetrate somewhat further due to the smaller absorption cross section $\sigma_{\rm bf,\nu} \propto \nu^{-3}$.  Figure \ref{fig:iondepth} shows the penetration depth $\Delta$ of photons with frequency $\nu \gtrsim \nu_{\rm thr}$, as calculated by integrating equation (\ref{eq:tau2}) for several characteristic values of the electron scattering optical depth $\tau_{\rm es} = \rho_{\rm ej}\kappa_{\rm es}\Delta_{\rm thr}$ through the ionized layer.  Appendix \ref{sec:iondepth} provides an analytic derivation of this result, from which the penetration depth is found to be
\be
\frac{\Delta}{\dtrh} = \left[ 4 - 3\left(\frac{\nu}{\nutrh}\right)^{-1} \right]^{q},
\ee
where $q=1$ if $\tau_{\rm es}\ll 1$ and $q=1/2$ if $\tau_{\rm es} \gg 1$.  Photons with $\nu \sim 2\nu_{\rm thr}$ thus penetrate to a distance which is $\approx 2-3$ times larger than those at the threshold frequency, while $\Delta$ increases much slower with frequency for $\nu \gg \nu_{\rm thr}$, due to the diminishing number of ionizing photons at higher frequency $E_{\rm nth,\nu} \propto \nu^{-1}$.
 
From equation (\ref{eq:dth}), the condition for ionization break-out (2$\Delta_{\rm thr} \gtrsim \Delta_{\rm sh}$) can be written as a condition on the scattering optical depth of the shocked ejecta $\tau_{\rm sh}^{\rm ej}$, which can be expressed in two limiting cases:
\begin{eqnarray}
&& \tau_{\rm es}^{\rm sh} < \tau_{\rm es,bo}^{\rm sh} = 
\left\{
\begin{array}{lr}
2\eta_{\rm thr}^{-1/2}
, &
\eta_{\rm thr} \ll 1 \\
2\eta_{\rm thr}^{-1},&
\eta_{\rm thr} \gg 1, \\
\end{array}
\right..
\label{eq:taues_bo}
\end{eqnarray}
where the pre-factor of $2$ accounts for the deeper penetration depth of photons with $\nu \gtrsim 2\nu_{\rm thr}$  (Fig.~\ref{fig:iondepth}), as characterizes a `typical' frequency in the band of interest.  

The break-out time corresponding to this optical depth is 
\begin{eqnarray}
t_{\rm bo} = \left(\frac{3M_{\rm ej}\kappa_{\rm es}}{4\pi v_{\rm ej}^{2}\tau_{\rm es,bo}^{\rm ej}}\right)^{1/2} \approx \nonumber 
\end{eqnarray}
\begin{eqnarray}
\left\{
\begin{array}{lr}
0.33\,{\rm yr}\,M_3^{3/4} v_9^{-5/4}T_{5}^{-0.2}\left(\frac{X_{A}}{0.1}\right)^{1/4}\left(\frac{L_{\rm sd}t}{10^{52}{\rm ergs}}\right)^{-1/4}Z_8^{3/4}
, 
(\eta_{\rm thr} \ll 1) \\
0.31\,{\rm yr}\,M_3 v_9^{-3/2}T_{5}^{-0.4}\left(\frac{X_{A}}{0.1}\right)^{1/2}\left(\frac{L_{\rm sd}t}{10^{52}{\rm ergs}}\right)^{-1/2}Z_8^{3/2},       
(\eta_{\rm thr} \gg 1), \\
\end{array}
\right.
\label{eq:tbo}
\end{eqnarray}
where we have adopted a fiducial value of the albedo $\mathcal{A}_{\nu} \sim 0.5$ and have used equation (\ref{eq:etaA}) for $\eta_{\rm thr}$ and equation (\ref{eq:tau_ej}) for $\tau^{\rm ej}_{\rm es}$ assuming $\delta = 0$.  For typical parameters ($M_3 \sim T_5 \sim 1$; $v_9 \sim 2$) and a millisecond pulsar with $L_{\rm sd}t \sim E_{\rm rot} \sim 10^{52}$ ergs, we have $t_{\rm bo} \sim 3$ weeks for helium ($Z = 2$; $X_{\rm He} = 0.15$), while $t_{\rm bo} \sim 2-3$ months for oxygen ($Z = 8$; $X_{\rm O} = 0.59$).  For iron ($Z = 26$; $X_{\rm Fe} = 0.05$), break-out occurs even later, $t_{\rm bo} \sim $ 4 months.  

Note the strong dependence of $t_{\rm bo}$ on the ejecta mass $M_{\rm ej}$.  If the final kinetic energy of the ejecta $M_{\rm ej}v_{\rm ej}^{2}/2$ derives mostly from the (approximately fixed) rotational energy of the pulsar $E_{\rm rot} \sim 10^{52}$ ergs, then $v_{\rm ej} \propto M_{\rm ej}^{-1/2}$ and hence $t_{\rm bo} \propto M_{\rm ej}^{11/8} (\propto M_{\rm ej}^{7/4})$ for $\eta_{\rm thr} \ll 1$ ($\eta_{\rm thr} \gg 1$).  Also note that the conditions for ionization break-out are in fact more complicated than suggested by equations (\ref{eq:taues_bo}) and (\ref{eq:tbo}) because the albedo $A_{\nu}$ and Compton temperature $T_{\rm C}^{Z,i}$ in the ionization front depend implicitly (and non-trivially) on the opacity ratio $\eta_{\rm thr}$.  For example, the Compton temperature generally increases with $Z$ due to the higher frequency photons allowed into the ionization layer, further weakening the dependence of $t_{\rm bo}$ on $Z$.

\section{Analytic solution for the photon penetration depth of a single atomic state}

\label{sec:iondepth}

The effective optical depth is defined by (eq.~[\ref{eq:tau2}])
\be
\int_0^{s} \rhoej \, \kabsnu(\spr) \, \left[1 + \tau(\spr)\right] \, d\spr = 1,
\label{eq:teff}
\ee
where
$\tau(\spr)=\kes\rhoej \spr$ and
\be
\rhoej\kabsnu(\spr) = \nA\sigmabf(\nu)\fn(\spr),
\label{eq:absbf}
\ee
where $n_A=\rho_{\rm sh}X_A/A m_p$ is the ion number density.
The neutral fraction is (Eq. [A4])
\be
\fn(s)\approx \left[ \frac{4\pi}{\arec\ne} \int_{\numin(s)} \frac{J_{\nu}}{h\nu} \, \sigmabf(\nu) \, d\nu \right]^{-1}.
\label{eq:ionfrac}
\ee
For any given frequency, Equation (\ref{eq:teff}) determines the depth to which the photons can penetrate before absorption,
or conversely the minimal energy of photons that can penetrate to depth $s$, i.e. $\numin(s)$.
However it depends on the ionization fraction (\ref{eq:ionfrac}), which in turn depends on $\numin(s)$.

To proceed, we substitute Equation (\ref{eq:absbf}) into (\ref{eq:teff}), divide both sides by $\nA\sigmabf(\nu)$ and differentiate, yielding
\be
\fn(s) \, \left[1 + \tau(s)\right] \, ds = \left[\frac{-1}{\nA\sigmabf(\nu)} \, \frac{d\ln{\sigmabf(\nu)}}{d\ln\nu} \, d\ln\nu \right]_{\nu=\numin}.
\label{eq:teff2}
\ee
At any given location $s$ the ionization fraction is an explicit function of $\numin$. 
Dividing Equation (\ref{eq:teff2}) by $\fn(\numin)$ separates the variables $s$ and $\numin$, yielding
\begin{eqnarray}
&&\left[1 + \tau(s)\right] \, ds \nonumber = \\
&& d\ln\numin\frac{-4\pi J_{\numin}}{\arec\ne\nA h} \, \frac{d\ln{\sigmabf(\numin)}}{d\ln\numin} \int_{\numin}^{\infty} \frac{J_{\nu}}{J_{\numin}} \, \frac{\sigmabf(\nu)}{\sigmabf(\numin)} \, d\ln\nu. \nonumber \\
\end{eqnarray}
For $J_{\nu}\propto\nu^{-1}$ and $\sigmabf(\nu)\propto \nu^{-3}$ we obtain
\be
\int_{\dtrh}^{\Delta} \left[1 + \tau(\spr)\right] \, d\spr = \frac{3\pi}{\arec\ne\nA h} \left[ J_{\nutrh} - J_{\nu} \right],
\ee
where $\dtrh$ is the penetration depth near the treshold frequency $\nutrh$.
It is determined by Equations (\ref{eq:teff}) -- (\ref{eq:ionfrac}) by setting
$\nu=\numin=\nutrh$, yielding
\be
\int_{0}^{\dtrh} \left[1 + \tau(\spr)\right] \, d\spr = \frac{\pi J_{\nutrh}}{\arec\ne\nA h}.
\ee
The penetration depth is in the limit that $\tau(\Delta_{\rm thr}) \ll 1$ or $\tau(\Delta_{\rm thr}) \gg 1$ is given by
\be
\frac{\Delta}{\dtrh} = \left[ 4 - 3\left(\frac{\nu}{\nutrh}\right)^{-1} \right]^{q},
\ee
where $q=1$ if $\tau(\Delta_{\rm thr}) \ll 1$ and $q=1/2$ if $\tau(\Delta_{\rm thr})\gg 1$.




\end{document}